\pdfoutput=1
\documentclass[12pt,a4paper]{article}

\usepackage{ifthen} 
\newboolean{pdflatex}
\setboolean{pdflatex}{true} 

\newboolean{articletitles}
\setboolean{articletitles}{true} 

\newboolean{uprightparticles}
\setboolean{uprightparticles}{false} 

\def\paperauthors{LHCb collaboration} 
\def\paperasciititle{Measurement of the lifetimes of promptly produced Omegac and Xic0 baryons} 
\def\papertitle{Measurement of the lifetimes\protect\\ of promptly produced\protect\\ $\mathchar"710A^{0}_{c}$ and $\mathchar"7104^{0}_{c}$ baryons} 
\def\paperkeywords{{High Energy Physics}, {LHCb}} 
\def\papercopyright{\the\year\ CERN for the benefit of the LHCb collaboration} 
\def\paperlicence{CC BY 4.0 licence}
\def\paperlicenceurl{https://creativecommons.org/licenses/by/4.0/}

\usepackage[top=1in, bottom=1.25in, left=1in, right=1in]{geometry}

\usepackage{microtype}
\usepackage{lineno}  
\usepackage{xspace} 
\usepackage{caption} 

\usepackage{graphicx}  
\usepackage{color}
\usepackage{colortbl}
\graphicspath{{./figs/}} 

\usepackage{amsmath} 
\usepackage{amssymb}
\usepackage{amsfonts}
\usepackage{upgreek} 

\newcommand*\patchAmsMathEnvironmentForLineno[1]{%
\expandafter\let\csname old#1\expandafter\endcsname\csname #1\endcsname
\expandafter\let\csname oldend#1\expandafter\endcsname\csname
end#1\endcsname
 \renewenvironment{#1}%
   {\linenomath\csname old#1\endcsname}%
   {\csname oldend#1\endcsname\endlinenomath}%
}
\newcommand*\patchBothAmsMathEnvironmentsForLineno[1]{%
  \patchAmsMathEnvironmentForLineno{#1}%
  \patchAmsMathEnvironmentForLineno{#1*}%
}
\AtBeginDocument{%
\patchBothAmsMathEnvironmentsForLineno{equation}%
\patchBothAmsMathEnvironmentsForLineno{align}%
\patchBothAmsMathEnvironmentsForLineno{flalign}%
\patchBothAmsMathEnvironmentsForLineno{alignat}%
\patchBothAmsMathEnvironmentsForLineno{gather}%
\patchBothAmsMathEnvironmentsForLineno{multline}%
\patchBothAmsMathEnvironmentsForLineno{eqnarray}%
}

\usepackage{hyperxmp}

\usepackage[pdftex,
            pdfauthor={\paperauthors},
            pdftitle={\paperasciititle},
            pdfkeywords={\paperkeywords},
            pdfcopyright={Copyright (C) \papercopyright},
            pdflicenseurl={\paperlicenceurl}]{hyperref}

\usepackage[colorinlistoftodos,textsize=scriptsize]{todonotes}

\usepackage[bottom,flushmargin,hang,multiple]{footmisc}

\usepackage[all]{hypcap} 

\usepackage{subcaption}

\usepackage{xspace} 
\usepackage{upgreek}

\def\lhcb   {\mbox{LHCb}\xspace}

\def\MagUp {\mbox{\em Mag\kern -0.05em Up}\xspace}

\ifthenelse{\boolean{uprightparticles}}%
{

 \def\Ppi         {\ensuremath{\uppi}\xspace}

 \def\PDelta      {\ensuremath{\Delta}\xspace}                 
 \def\PXi         {\ensuremath{\Xi}\xspace}                 
 \def\PLambda     {\ensuremath{\Lambda}\xspace}                 
 \def\PSigma      {\ensuremath{\Sigma}\xspace}                 
 \def\POmega      {\ensuremath{\Omega}\xspace}                 
 \def\PUpsilon    {\ensuremath{\Upsilon}\xspace}

 \def\PB      {\ensuremath{\mathrm{B}}\xspace}                 
                  
 \def\PD      {\ensuremath{\mathrm{D}}\xspace}

 \def\PK      {\ensuremath{\mathrm{K}}\xspace}

 \def\Pb      {\ensuremath{\mathrm{b}}\xspace}                 
 \def\Pc      {\ensuremath{\mathrm{c}}\xspace}

 \def\Pi      {\ensuremath{\mathrm{i}}\xspace}

 \def\Pp      {\ensuremath{\mathrm{p}}\xspace}

 \def\Ps      {\ensuremath{\mathrm{s}}\xspace}

 \def\thebaroffset{0.0em}
}
{

 \def\Ppi         {\ensuremath{\pi}\xspace}

 \mathchardef\PDelta="7101
 \mathchardef\PXi="7104
 \mathchardef\PLambda="7103
 \mathchardef\PSigma="7106
 \mathchardef\POmega="710A
 \mathchardef\PUpsilon="7107
                  
 \def\PB      {\ensuremath{B}\xspace}                 
                  
 \def\PD      {\ensuremath{D}\xspace}

 \def\PK      {\ensuremath{K}\xspace}

 \def\Pb      {\ensuremath{b}\xspace}                 
 \def\Pc      {\ensuremath{c}\xspace}

 \def\Pi      {\ensuremath{i}\xspace}

 \def\Pp      {\ensuremath{p}\xspace}

 \def\Ps      {\ensuremath{s}\xspace}

 \def\thebaroffset{0.18em}
}
\newcommand{\offsetoverline}[2][\thebaroffset]{\kern #1\overline{\kern -#1 #2}}%

\makeatletter
\ifcase \@ptsize \relax
  \newcommand{\miniscule}{\@setfontsize\miniscule{4}{5}}
\or
  \newcommand{\miniscule}{\@setfontsize\miniscule{5}{6}}
\or
  \newcommand{\miniscule}{\@setfontsize\miniscule{5}{6}}
\fi
\makeatother

\DeclareRobustCommand{\optbar}[1]{\shortstack{{\miniscule (\rule[.5ex]{1.25em}{.18mm})}
  \\ [-.7ex] $#1$}}

\def\squark    {{\ensuremath{\Ps}}\xspace}

\def\cquark    {{\ensuremath{\Pc}}\xspace}

\def\bquark    {{\ensuremath{\Pb}}\xspace}

\def\pion   {{\ensuremath{\Ppi}}\xspace}

\def\pip    {{\ensuremath{\pion^+}}\xspace}
\def\pim    {{\ensuremath{\pion^-}}\xspace}

\def\kaon    {{\ensuremath{\PK}}\xspace}

\def\KorKbar {\kern \thebaroffset\optbar{\kern -\thebaroffset \PK}{}\xspace}

\def\Kp      {{\ensuremath{\kaon^+}}\xspace}
\def\Km      {{\ensuremath{\kaon^-}}\xspace}

\def\Dbar    {{\ensuremath{\offsetoverline{\PD}}}\xspace}
\def\D       {{\ensuremath{\PD}}\xspace}

\def\DorDbar {\kern \thebaroffset\optbar{\kern -\thebaroffset \PD}\xspace}
\def\Dz      {{\ensuremath{\D^0}}\xspace}
\def\Dzb     {{\ensuremath{\Dbar{}^0}}\xspace}
\def\Dp      {{\ensuremath{\D^+}}\xspace}
\def\Dm      {{\ensuremath{\D^-}}\xspace}

\def\DpDm    {\ensuremath{\Dp {\kern -0.16em \Dm}}\xspace}

\def\B       {{\ensuremath{\PB}}\xspace}

\def\BorBbar {\kern \thebaroffset\optbar{\kern -\thebaroffset \PB}\xspace}

\def\Bd      {{\ensuremath{\B^0}}\xspace}

\def\BdorBdbar {\kern \thebaroffset\optbar{\kern -\thebaroffset \Bd}\xspace}

\def\Bs      {{\ensuremath{\B^0_\squark}}\xspace}

\def\BsorBsbar {\kern \thebaroffset\optbar{\kern -\thebaroffset \Bs}\xspace}

\def\Y#1S{\ensuremath{\PUpsilon{(#1S)}}\xspace}

\def\proton      {{\ensuremath{\Pp}}\xspace}

\def\Lz          {{\ensuremath{\PLambda}}\xspace}

\def\LorLbar     {\kern \thebaroffset\optbar{\kern -\thebaroffset \PLambda}\xspace}

\def\Xires       {{\ensuremath{\PXi}}\xspace}

\def\Omegares    {{\ensuremath{\POmega}}\xspace}

\def\Lc          {{\ensuremath{\Lz^+_\cquark}}\xspace}

\def\Xicz        {{\ensuremath{\Xires^0_\cquark}}\xspace}
\def\Xicp        {{\ensuremath{\Xires^+_\cquark}}\xspace}

\def\Omegac      {{\ensuremath{\Omegares^0_\cquark}}\xspace}

\def\Lb           {{\ensuremath{\Lz^0_\bquark}}\xspace}

\def\to                 {\ensuremath{\rightarrow}\xspace}

\def\AT#1     {\ensuremath{A_{\mathrm{T}}^{#1}}\xspace}           

\def\C#1      {\ensuremath{\mathcal{C}_{#1}}\xspace}                       
\def\Cp#1     {\ensuremath{\mathcal{C}_{#1}^{'}}\xspace}                    
\def\Ceff#1   {\ensuremath{\mathcal{C}_{#1}^{\mathrm{(eff)}}}\xspace}        
\def\Cpeff#1  {\ensuremath{\mathcal{C}_{#1}^{'\mathrm{(eff)}}}\xspace}       
\def\Ope#1    {\ensuremath{\mathcal{O}_{#1}}\xspace}                       
\def\Opep#1   {\ensuremath{\mathcal{O}_{#1}^{'}}\xspace}                    

\newcommand{\nospaceunit}[1]{\ensuremath{\text{#1}}}       
\newcommand{\aunit}[1]{\ensuremath{\text{\,#1}}}       

\newcommand{\tev}{\aunit{Te\kern -0.1em V}\xspace}
\newcommand{\gev}{\aunit{Ge\kern -0.1em V}\xspace}
\newcommand{\mev}{\aunit{Me\kern -0.1em V}\xspace}
\newcommand{\kev}{\aunit{ke\kern -0.1em V}\xspace}
\newcommand{\ev}{\aunit{e\kern -0.1em V}\xspace}
 
\newcommand{\mevc}{\ensuremath{\aunit{Me\kern -0.1em V\!/}c}\xspace}
\newcommand{\gevc}{\ensuremath{\aunit{Ge\kern -0.1em V\!/}c}\xspace}
\newcommand{\mevcc}{\ensuremath{\aunit{Me\kern -0.1em V\!/}c^2}\xspace}
\newcommand{\gevcc}{\ensuremath{\aunit{Ge\kern -0.1em V\!/}c^2}\xspace}

\def\mum  {\ensuremath{\,\upmu\nospaceunit{m}}\xspace}

\def\fb   {\ensuremath{\aunit{fb}}\xspace}
\def\invfb   {\ensuremath{\fb^{-1}}\xspace}

\def\ps   {\ensuremath{\aunit{ps}}\xspace}
\def\fs   {\aunit{fs}}

\newcommand{\chisq}{\ensuremath{\chi^2}\xspace}
\newcommand{\chisqndf}{\ensuremath{\chi^2/\mathrm{ndf}}\xspace}
\newcommand{\chisqip}{\ensuremath{\chi^2_{\text{IP}}}\xspace}

\def\gsim{{~\raise.15em\hbox{$>$}\kern-.85em
          \lower.35em\hbox{$\sim$}~}\xspace}
\def\lsim{{~\raise.15em\hbox{$<$}\kern-.85em
          \lower.35em\hbox{$\sim$}~}\xspace}

\def\sPlot{\mbox{\em sPlot}\xspace}

\def\pt         {\ensuremath{p_{\mathrm{T}}}\xspace}

\def\ptot       {\ensuremath{p}\xspace}

\def\mrad{\aunit{mrad}\xspace}

\def\evtgen     {\mbox{\textsc{EvtGen}}\xspace}

\def\geant      {\mbox{\textsc{Geant4}}\xspace}

\def\photos     {\mbox{\textsc{Photos}}\xspace}

\def\pythia     {\mbox{\textsc{Pythia}}\xspace}

\def\tell1  {TELL1\xspace}
\def\ukl1   {UKL1\xspace}

\newcommand{\ie}{\mbox{\itshape i.e.}\xspace}

\newcommand{\pkkpi}{\ensuremath{\proton\Km\Km\pip}\xspace}
\newcommand{\kkpipi}{\ensuremath{\Kp\Km\pip\pim}\xspace}

\newcommand{\xiczval}{148.0}
\newcommand{\xiczstat}{2.3}
\newcommand{\xiczsys}{2.2}
\newcommand{\xiczext}{0.2}

\newcommand{\combxiczval}{152.0}
\newcommand{\combxiczerr}{2.0}

\newcommand{\omegacval}{276.5}
\newcommand{\omegacstat}{13.4}
\newcommand{\omegacsys}{4.4}
\newcommand{\omegacext}{0.7}

\newcommand{\combomegacval}{274.5}
\newcommand{\combomegacerr}{12.4}

\def\logchisqip {\ensuremath{\log_{10}\chisqip}\xspace}
\mathchardef\mhyphen="2D

\newcommand{\xx}{\ensuremath{\kern 0.5em }}

\def\sPlot{\mbox{\em sPlot}\xspace}

\usepackage{multirow} 
\usepackage{booktabs} 
\usepackage{rotating}

\usepackage{cite} 
\usepackage{mciteplus}

\usepackage{longtable} 

\begin{document}

\renewcommand{\thefootnote}{\fnsymbol{footnote}}
\setcounter{footnote}{1}

\begin{titlepage}
\pagenumbering{roman}

\vspace*{-1.5cm}
\centerline{\large EUROPEAN ORGANIZATION FOR NUCLEAR RESEARCH (CERN)}
\vspace*{1.5cm}
\noindent
\begin{tabular*}{\linewidth}{lc@{\extracolsep{\fill}}r@{\extracolsep{0pt}}}
\ifthenelse{\boolean{pdflatex}}
{\vspace*{-1.5cm}\mbox{\!\!\!\includegraphics[width=.14\textwidth]{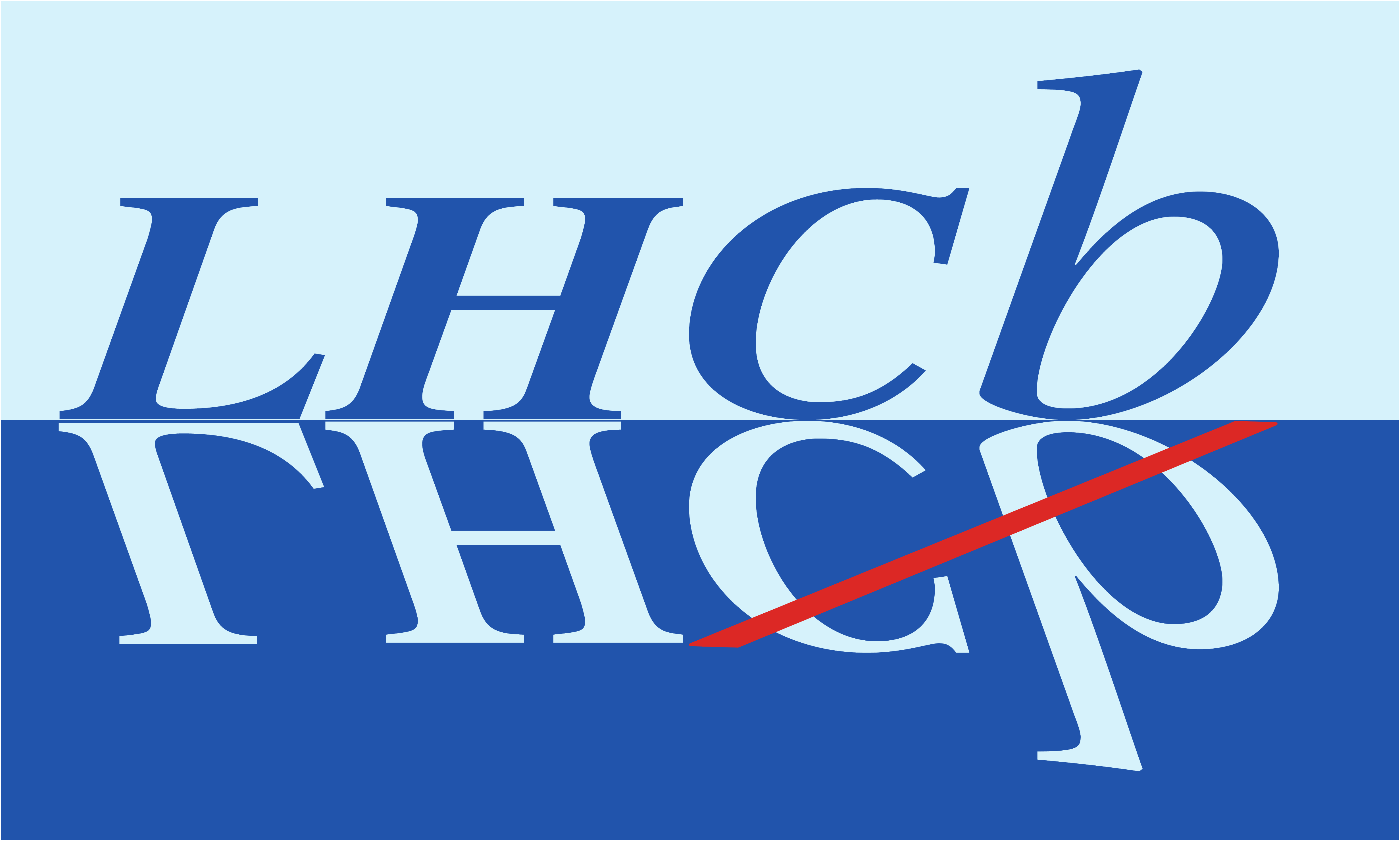}} & &}%
{\vspace*{-1.2cm}\mbox{\!\!\!\includegraphics[width=.12\textwidth]{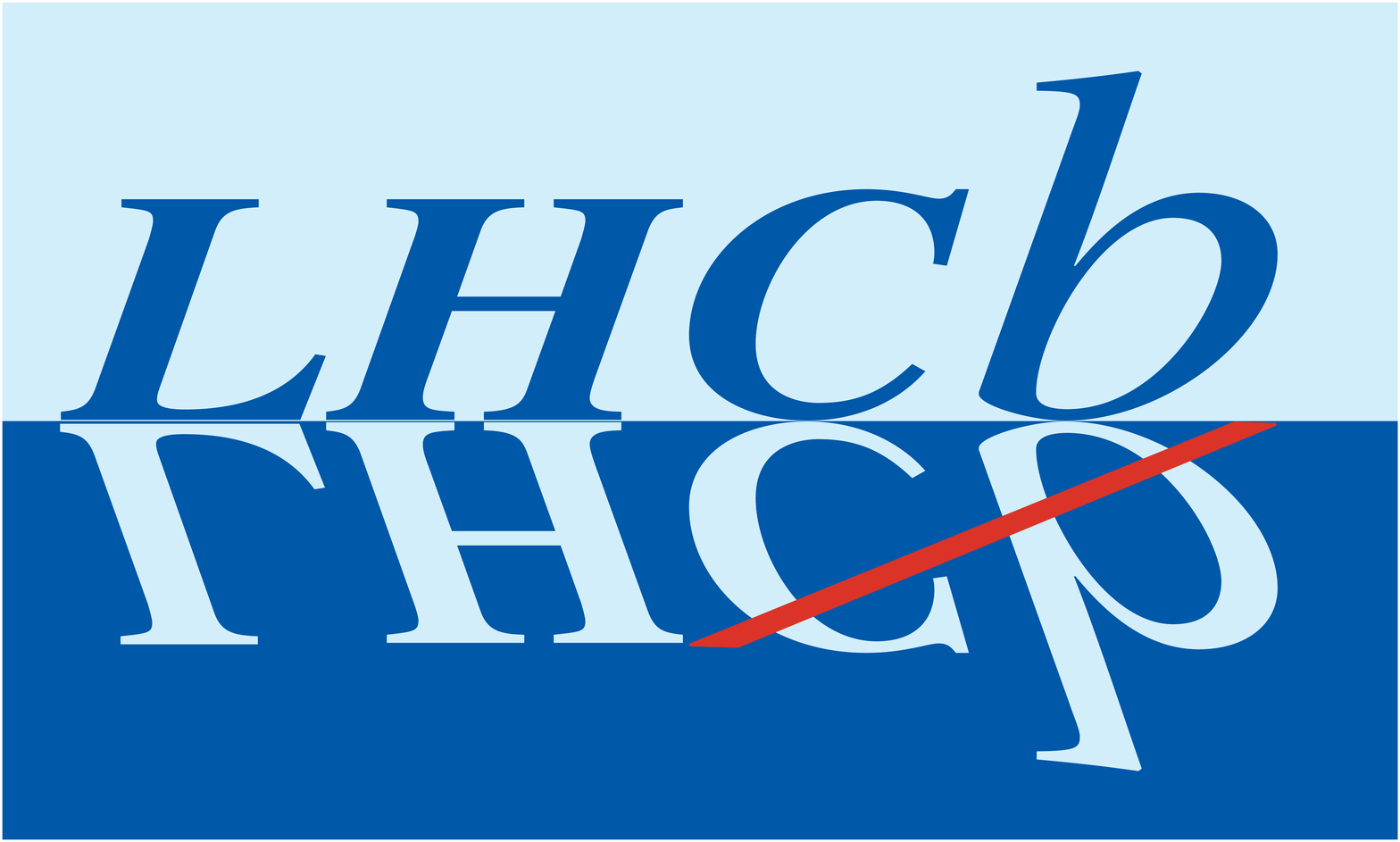}} & &}%
\\
 & & CERN-EP-2021-167 \\  
 & & LHCb-PAPER-2021-021 \\  
 & & February 19, 2021 \\ 
 & & \\
\end{tabular*}

\vspace*{4.0cm}

{\normalfont\bfseries\boldmath\huge
\begin{center}
  \papertitle 
\end{center}
}

\vspace*{2.0cm}

\begin{center}
\paperauthors\footnote{Authors are listed at the end of this paper.}
\end{center}

\vspace{\fill}

\begin{abstract}
  \noindent
  A measurement of the lifetimes of the $\mathchar"710A^{0}_{c}$ and $\mathchar"7104^{0}_{c}$ baryons is reported using proton--proton collision data at a centre-of-mass energy of $13\text{\,Te\kern -0.1em V}$, corresponding to an integrated luminosity of $5.4\text{\,fb}^{-1}$ collected by the LHCb experiment. The $\mathchar"710A^{0}_{c}$ and $\mathchar"7104^{0}_{c}$ baryons are produced directly from proton interactions and reconstructed in the $pK^{-}K^{-}\pi^{+}$ final state. The $\mathchar"710A^{0}_{c}$ lifetime is measured to be \mbox{$276.5\pm13.4\pm4.4\pm0.7\text{\,fs}$}, and the $\mathchar"7104^{0}_{c}$ lifetime is measured to be \mbox{$148.0\pm2.3\pm2.2\pm0.2\text{\,fs}$}, where the first uncertainty is statistical, the second systematic, and the third due to the uncertainty on the $D^{0}$ lifetime. These results confirm previous LHCb measurements based on semileptonic beauty-hadron decays, which disagree with earlier results of a four times shorter $\mathchar"710A^{0}_{c}$ lifetime, and provide the single most precise measurement of the $\mathchar"710A^{0}_{c}$ lifetime.

\end{abstract}

\vspace*{1.0cm}

\begin{center}
  Published in Science Bulletin 2022, 67(5): 479-487
\end{center}

\vspace{\fill}

{\footnotesize 
\centerline{\copyright~\papercopyright. \href{\paperlicenceurl}{\paperlicence}.}}
\vspace*{2mm}

\end{titlepage}


\newpage
\setcounter{page}{2}
\mbox{~}
%
%


\renewcommand{\thefootnote}{\arabic{footnote}}
\setcounter{footnote}{0}

\cleardoublepage


\pagestyle{plain} 
\setcounter{page}{1}
\pagenumbering{arabic}

\section{Introduction}
The \lhcb collaboration has previously measured the lifetimes of \Omegac and \Xicz baryons,
with valence quark content of $ssc$ and $dsc$, respectively,
using candidates from semileptonic beauty-hadron decays~\cite{LHCb-PAPER-2018-028,LHCb-PAPER-2019-008}.
The measured \Omegac lifetime, $\tau_{\Omegac}$, is nearly four times larger than the previous world average~\cite{PDG2018},
which is inconsistent at a level of seven standard deviations.
The measured \Xicz lifetime, $\tau_{\Xicz}$, is larger than the previous world average by three standard deviations.
Resolving these discrepancies is important for theoretical calculations.
Lifetime measurements of hadrons containing heavy \mbox{quarks $Q$,} \ie charm or beauty quarks, 
provide input needed to test precisely the Standard Model and search for physics beyond. 

Heavy quark expansion~\cite{Khoze:1983yp,Bigi:1991ir,Bigi:1992su,Blok:1992hw,*Blok:1992he,
Neubert:1997gu,Uraltsev:1998bk,Bigi:1995jr}
is an effective theory used to calculate
the lifetimes of these hadrons through an expansion in inverse powers of the mass of the heavy quark, $m_{Q}$.
The lowest-order term in the expansion depends only on $m_{Q}$ and  
contributes equally to the decay width of all hadrons with the same heavy quark.
Differences in predicted lifetimes are expected to arise from higher-order effects,
such as weak $W$-annihilation and Pauli interference,
due to the presence of different spectator quarks. 
Measurements of charmed-hadron lifetimes are particularly sensitive to these higher-order contributions,
as these corrections typically increase as $m_Q$
decreases and are large for charmed hadrons~\cite{Kirk:2017juj,Cheng:2015iom,Lenz:2013aua,Bianco:2003vb,Bellini:1996ra,Blok:1991st}.
A lifetime hierarchy of \mbox{$\tau_{\Xicp}>\tau_{\Lc}>\tau_{\Xicz}>\tau_{\Omegac}$}
 is obtained considering higher-order effects~\cite{Blok:1991st,Bellini:1996ra,Cheng:1997xba,Bianco:2003vb,Cheng:2015iom},
whereas the ordering of \mbox{$\tau_{\Xicp}>\tau_{\Omegac}>\tau_{\Lc}>\tau_{\Xicz}$}
can be obtained depending on the treatment of even higher-order effects~\cite{Blok:1991st,Cheng:2018rkz,cheng2021charmed}.
Knowledge on the lifetimes is also required to make comparisons between measured
branching fractions of charmed baryons and corresponding theoretical predictions
for their partial decay widths.

The \lhcb experiment has recorded an unprecedented number of charmed baryons,
produced both at the primary proton--proton ($pp$) collision vertex (PV), referred to as prompt,
and from decays of beauty hadrons.
In this paper, a measurement of the lifetimes of \Omegac and \Xicz baryons is reported
based on a sample of \Omegac and \Xicz baryons promptly produced in $pp$ collisions 
at a centre-of-mass energy of $13\tev$. 
The data sample was collected by the \lhcb experiment between 2016--2018
and corresponds to an integrated luminosity of $5.4\invfb$.
The \Omegac and \Xicz baryons are reconstructed in the $\pkkpi$ final state.\footnote{The inclusion of charge-conjugate processes is implied throughout.}
In order to avoid experimenter’s bias, the results of the analysis were not examined until the full procedure had been finalised.
Prompt $\Dz\to\kkpipi$ decays are used as a control mode
in order to reduce systematic uncertainties and to validate the analysis procedure.

\section{Detector and simulation}
The \lhcb detector~\cite{LHCb-DP-2008-001,LHCb-DP-2014-002} is a single-arm forward
spectrometer covering the \mbox{pseudorapidity} range $2<\eta <5$,
designed for the study of particles containing \bquark or \cquark quarks.
The tracking system provides a measurement of the momentum, \ptot, of charged particles with
a relative uncertainty that varies from 0.5\% at low momentum to 1.0\% at 200\gevc.
The minimum distance of a track to a
PV, the impact parameter (IP),
is measured with a resolution of $(15+29/\pt)\mum$,
where \pt is the component of the momentum transverse to the beam, in\,\gevc.
Different types of charged hadrons are distinguished using information
from two ring-imaging Cherenkov detectors.
Photons, electrons and hadrons are identified by a calorimeter system consisting of
scintillating-pad and preshower detectors, an electromagnetic and a hadronic calorimeter.
The online event selection is performed by a trigger,
which consists of a hardware stage, based on information from the calorimeter and muon
systems, followed by a two-level software stage, which applies a full event reconstruction.
Between the two software stages, an alignment and calibration of the detector
is performed in near real-time and their results are used in the trigger.
The same alignment and calibration information is propagated to the offline reconstruction,
ensuring consistent and high-quality particle identification (PID) information
between the trigger and offline software.
The identical performance of the online and offline
reconstruction offers the opportunity to perform physics analyses
directly using candidates reconstructed in the trigger
\cite{LHCb-DP-2012-004,LHCb-DP-2016-001},
which this analysis utilises.

Simulated samples are used to model the effects of the detector acceptance and the
imposed selection requirements, and to study the modelling of the discriminating
variables between signal and background candidates.
In the simulation, $pp$ collisions are generated using
\pythia~\cite{Sjostrand:2006za,Sjostrand:2007gs} 
with a specific \lhcb configuration~\cite{LHCb-PROC-2010-056}.
Decays of unstable particles are described by \evtgen~\cite{Lange:2001uf},
in which final-state radiation is generated using \photos~\cite{davidson2015photos}.
The interaction of the generated particles with the detector, and its response,
are implemented using the \geant toolkit~\cite{Allison:2006ve, *Agostinelli:2002hh}
as described in Ref.~\cite{LHCb-PROC-2011-006}.
The underlying $pp$ interaction is reused multiple times,
with an independently generated signal decay for each~\cite{LHCb-DP-2018-004}.

\section{Candidate selection}
Candidate decays of  charmed hadrons are reconstructed
through the \mbox{$\Omegac\to\pkkpi$}, \mbox{$\Xicz\to\pkkpi$}, and \mbox{$\Dz\to\kkpipi$} decay modes.
All final-state charged particle candidates are required to be inconsistent with originating from any PV.
The PV associated to a single charged particle is defined to be the PV with the smallest \chisqip,
where \chisqip is defined as the difference in the vertex-fit \chisq of a given PV
reconstructed with and without the given particle.
Each of the final-state particles is required to have good track quality,
large transverse and total momentum,
and particle-identification information consistent with the corresponding
\proton, \kaon, or \pion hypothesis.
The angle between each pair of final-state particles is required to be larger than $0.5\mrad$
to avoid selecting duplicate tracks.
The charmed hadron candidates are required to have a decay vertex with good quality
that is displaced from its associated PV.
The angle between the reconstructed momentum vector of a charmed baryon candidate and the direction from its associated PV to its decay vertex, the direction angle,
is required to be small to suppress combinatorial background.

To improve further the signal purity,
a multivariate classifier is trained
based on the adaptive boosted decision tree~(BDT) algorithm~\cite{Breiman,AdaBoost}
implemented in the TMVA toolkit~\cite{Hocker:2007ht,*TMVA4}.
The classifier is trained using simulated prompt $\Omegac\to\pkkpi$ decays as signal,
and data from the \Omegac mass sidebands of $25<|m(\pkkpi)-m_{\Omegac}|<75\mevcc$
as background.
Here, $m(\pkkpi)$ is the invariant mass of the decay products of the charmed baryon candidate and $m_{\Omegac}$ is the known \Omegac mass~\cite{PDG2020}.
The sideband ranges from $5$ to $14$ times the invariant-mass resolution.
The lifetime of the simulated \Omegac decays is $250\fs$.
Eleven input variables are used in the training,
including: the \chisq of the \Omegac decay-vertex fit;
the transverse momentum, pseudorapidity and the direction angle of the \Omegac candidate;
the transverse momenta as well as the minimal transverse momentum of the four final-state particles;
and the natural logarithm of the sum and minimum \chisqip of the four final-state particles.
A requirement on the BDT response is chosen which selects approximately $99\%$ of \Omegac signal decays
while rejecting about $60\%$ of the background.
The same requirement is also applied to the \Xicz signal decays.
The number of signal candidates for the \Omegac, \Xicz, and \Dz decay modes are determined using the $[2645,2745]\mevcc$, $[2421,2521]\mevcc$, and $[1835,1915]\mevcc$ invariant-mass regions, respectively.

Specific trigger requirements are applied to candidates
to ensure a precise estimation of the selection efficiency as a function of decay time.
In the offline selection, trigger signals are associated with reconstructed particles.
Selection requirements can therefore be made on the trigger selection itself
and if the decision was due to the signal candidate.
At the hardware stage, at least one of the final-state tracks is required to 
deposit large transverse energy in the hadronic calorimeter.
At the software stage, at least one of the final-state tracks is required to 
pass a MatrixNet classifier~\cite{matrix_net},
which is trained to select displaced tracks~\cite{LHCb-DP-2019-001}.


\section{Prompt yield determination}
Charmed hadron candidates are split into intervals of their decay time,
which is calculated using the PV, its decay vertex, and its measured momentum.
The signal yields are then determined in each interval.
The interval boundaries of the \Omegac sample
are chosen to have a similar yield of \Omegac signals in each interval,
and correspond to 
$[0.45, 0.52, 0.57, 0.63, 0.69, 0.75, 0.81, 0.90, 1.05, 2.00]\ps$.
For the \Xicz sample, the same boundaries are used except the last interval,
which, for computational simplicity, is not included as the yield is consistent with zero.
The same boundaries are used for the \Dz control mode as for the signal modes.
Two variables are used to discriminate signal decays from different background contributions.
One is the invariant mass of the charmed hadron,
which is used to distinguish decays from combinatorial background
due to the random combinations of tracks.
The other is the logarithm of the \chisqip of the charmed hadron, $\logchisqip$,
which is used to separate prompt candidates from those 
produced in decays of beauty hadrons.
The $\logchisqip$ distribution for signal decays
has smaller mean values than for those originating from beauty-hadron decays
due to the lifetime of the ancestor beauty hadron.

\begin{figure}[tb!]
  \centering
  \begin{subfigure}[t]{0.45\textwidth}
  \includegraphics[width=\linewidth]{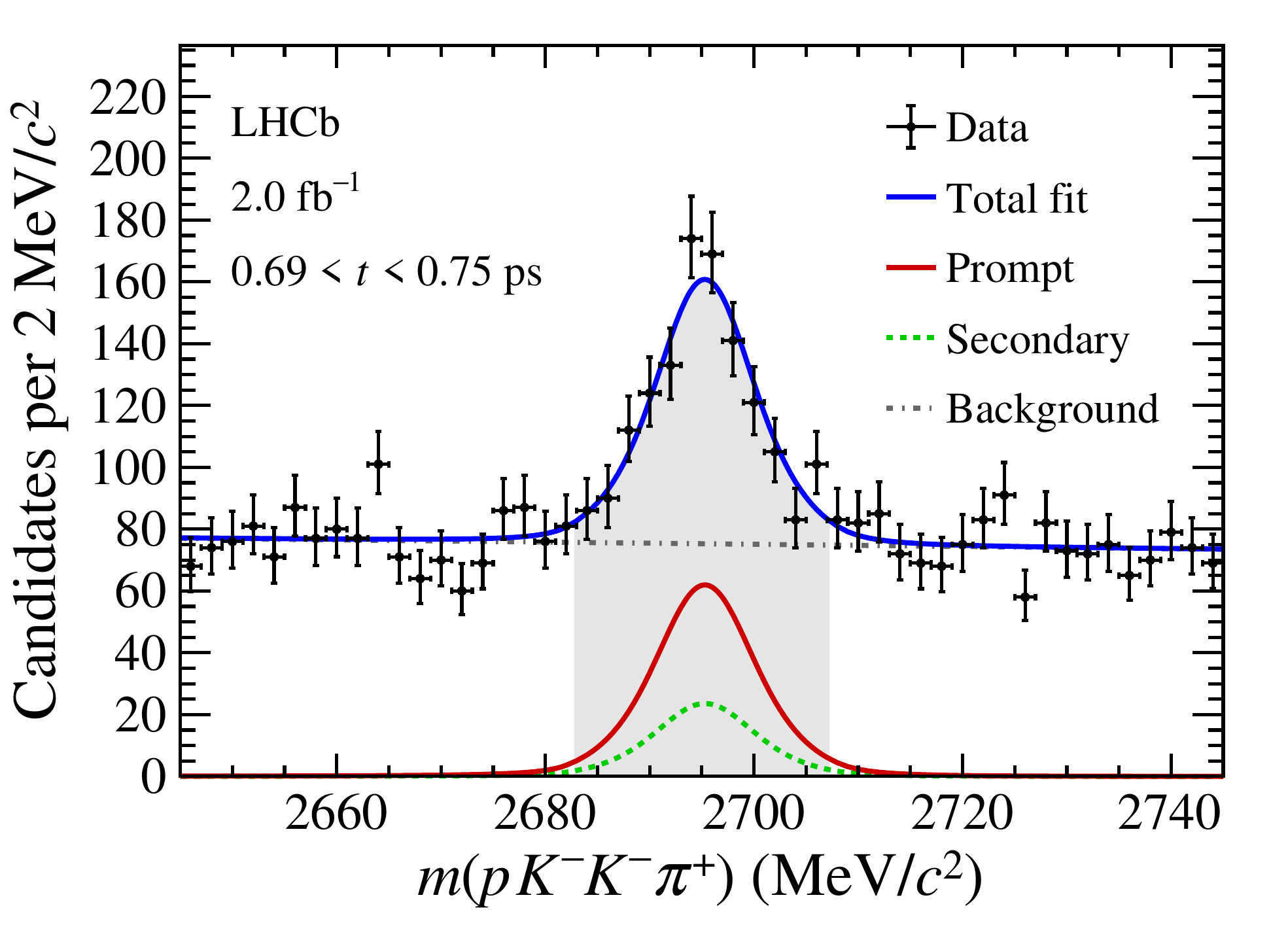}
  \caption{}
  \end{subfigure}
  \begin{subfigure}[t]{0.45\textwidth}
  \includegraphics[width=\linewidth]{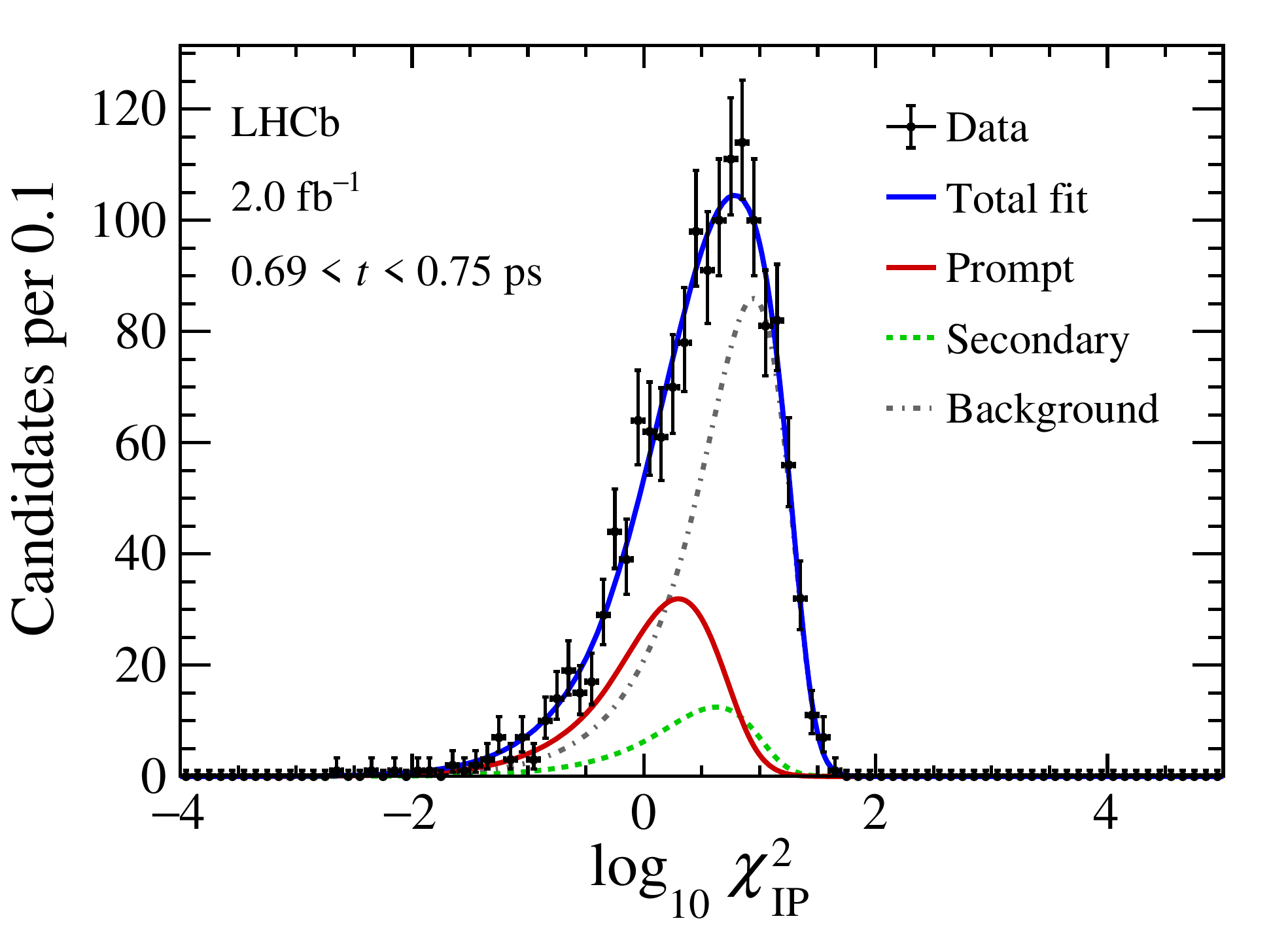}
  \caption{}
  \end{subfigure}
  \begin{subfigure}[t]{0.45\textwidth}
  \includegraphics[width=\linewidth]{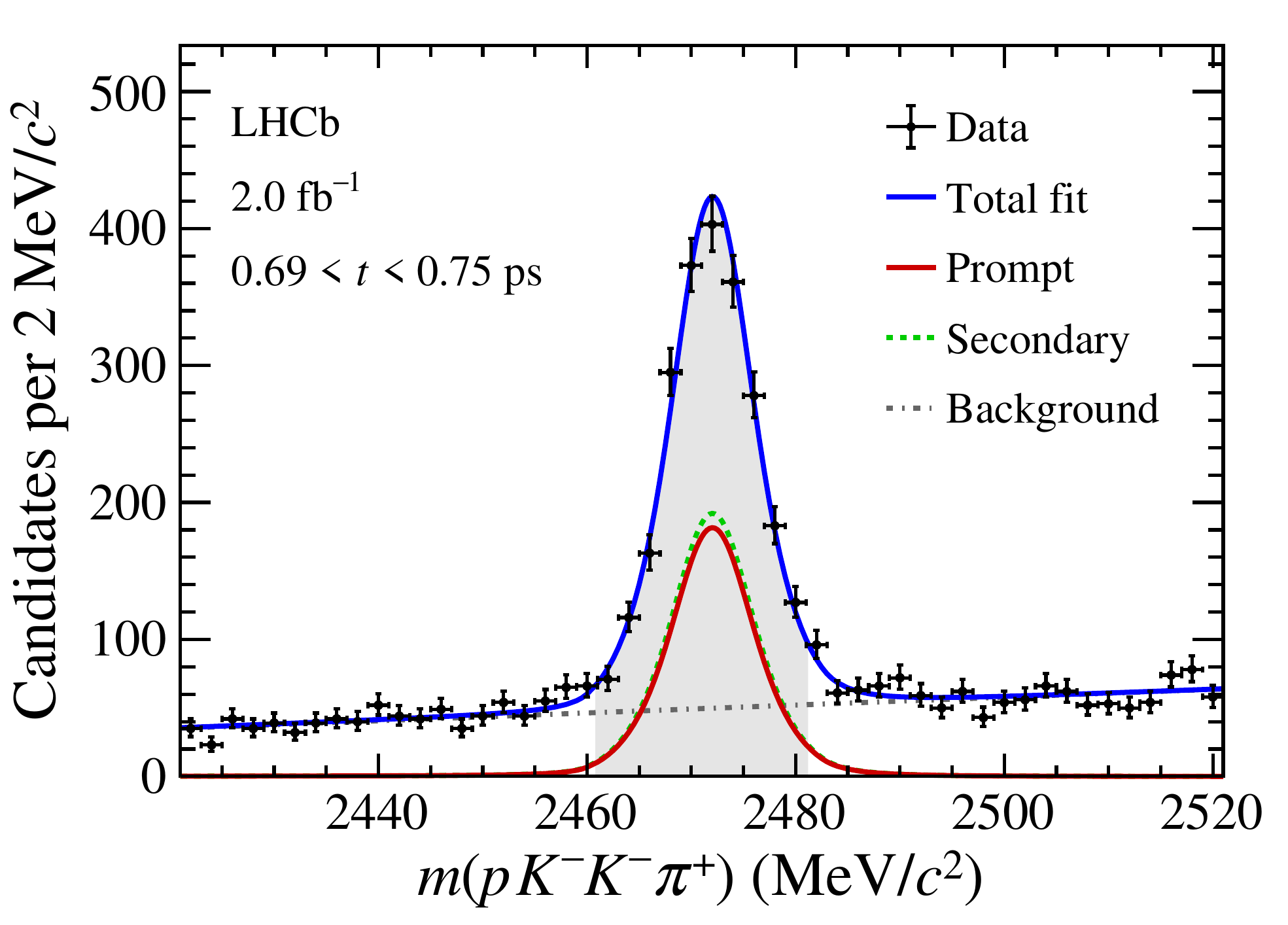}
  \caption{}
  \end{subfigure}
  \begin{subfigure}[t]{0.45\textwidth}
  \includegraphics[width=\linewidth]{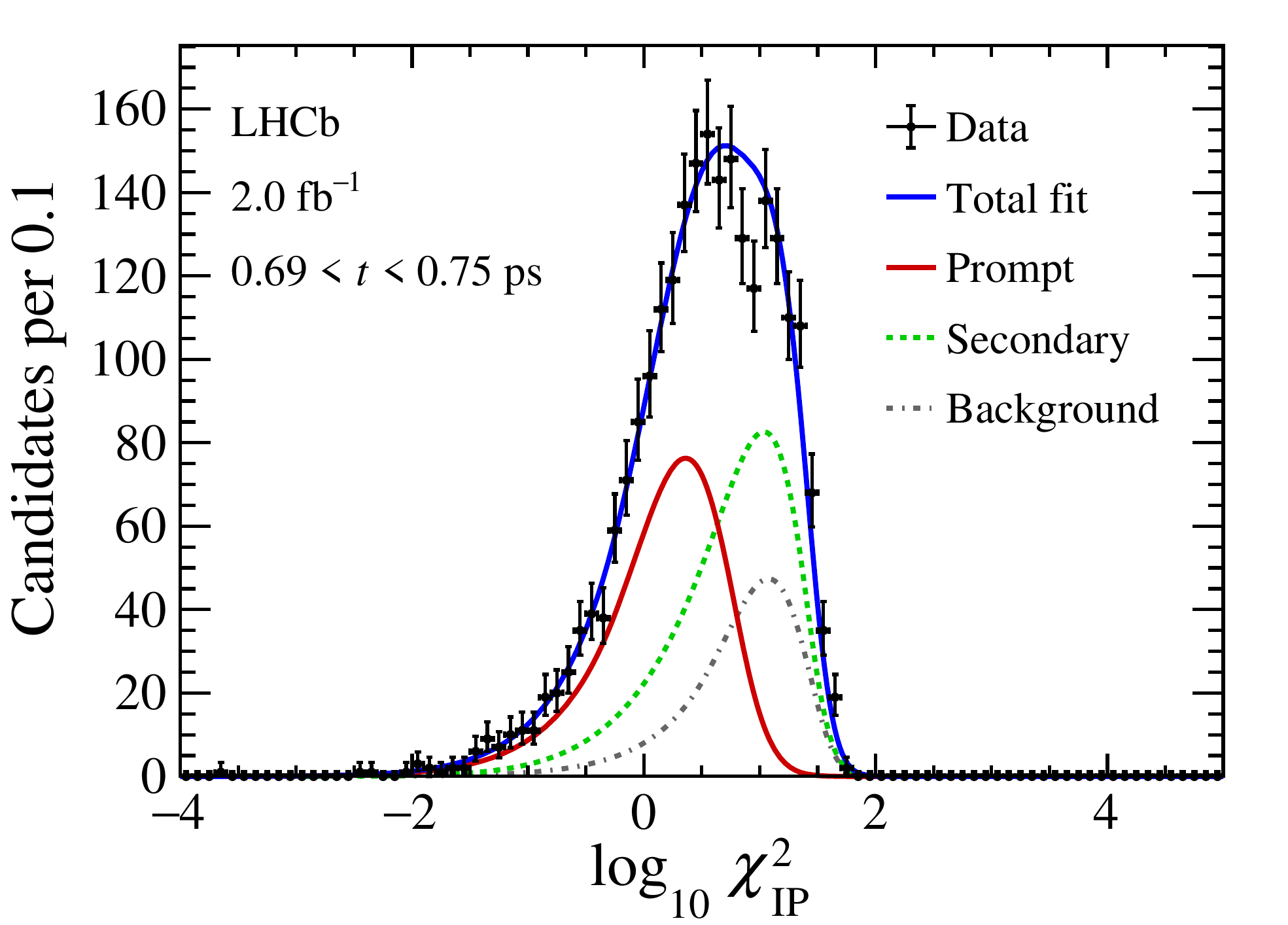}
  \caption{}
  \end{subfigure}
  \begin{subfigure}[t]{0.45\textwidth}
  \includegraphics[width=\linewidth]{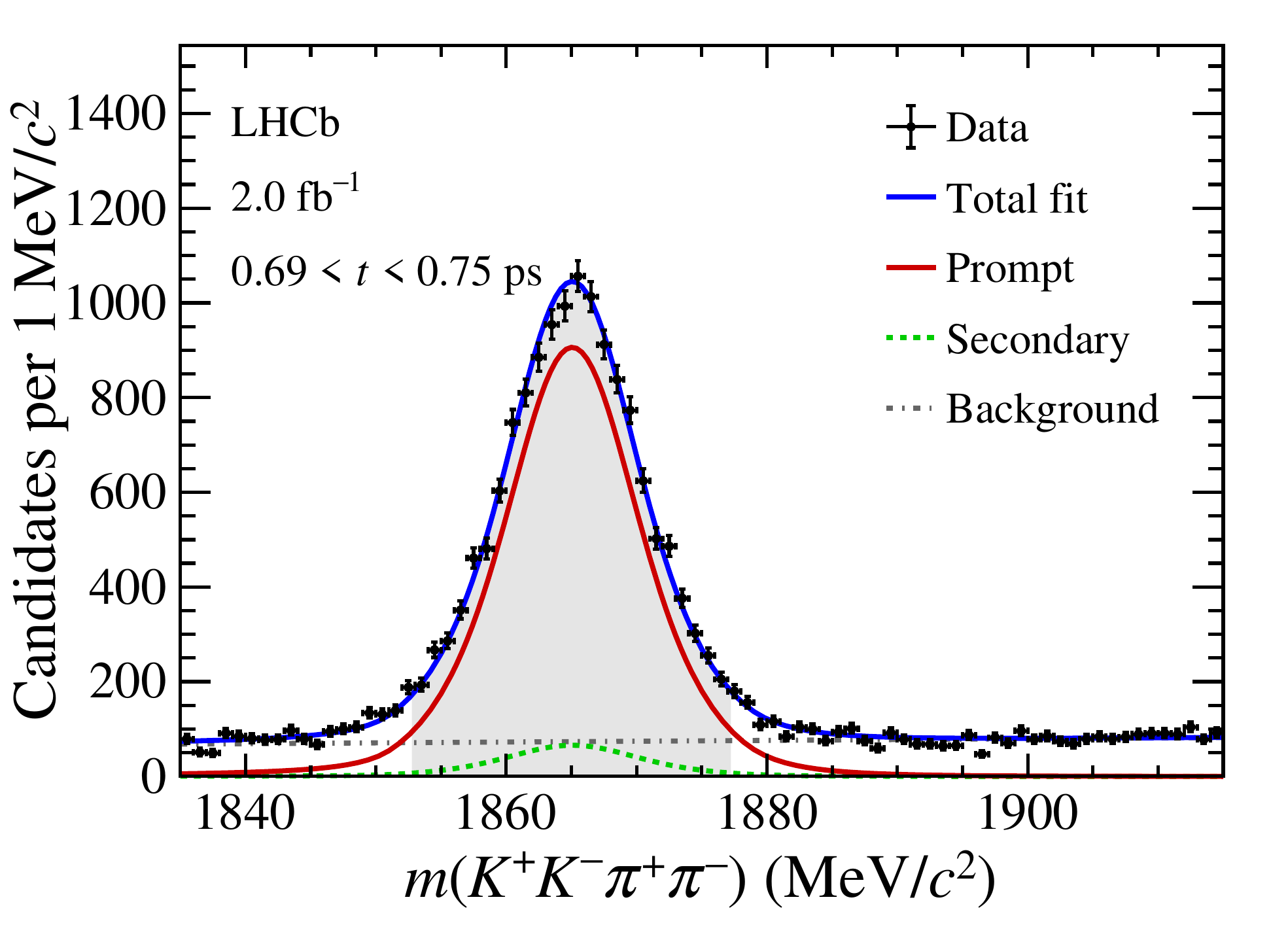}
  \caption{}
  \end{subfigure}
  \begin{subfigure}[t]{0.45\textwidth}
  \includegraphics[width=\linewidth]{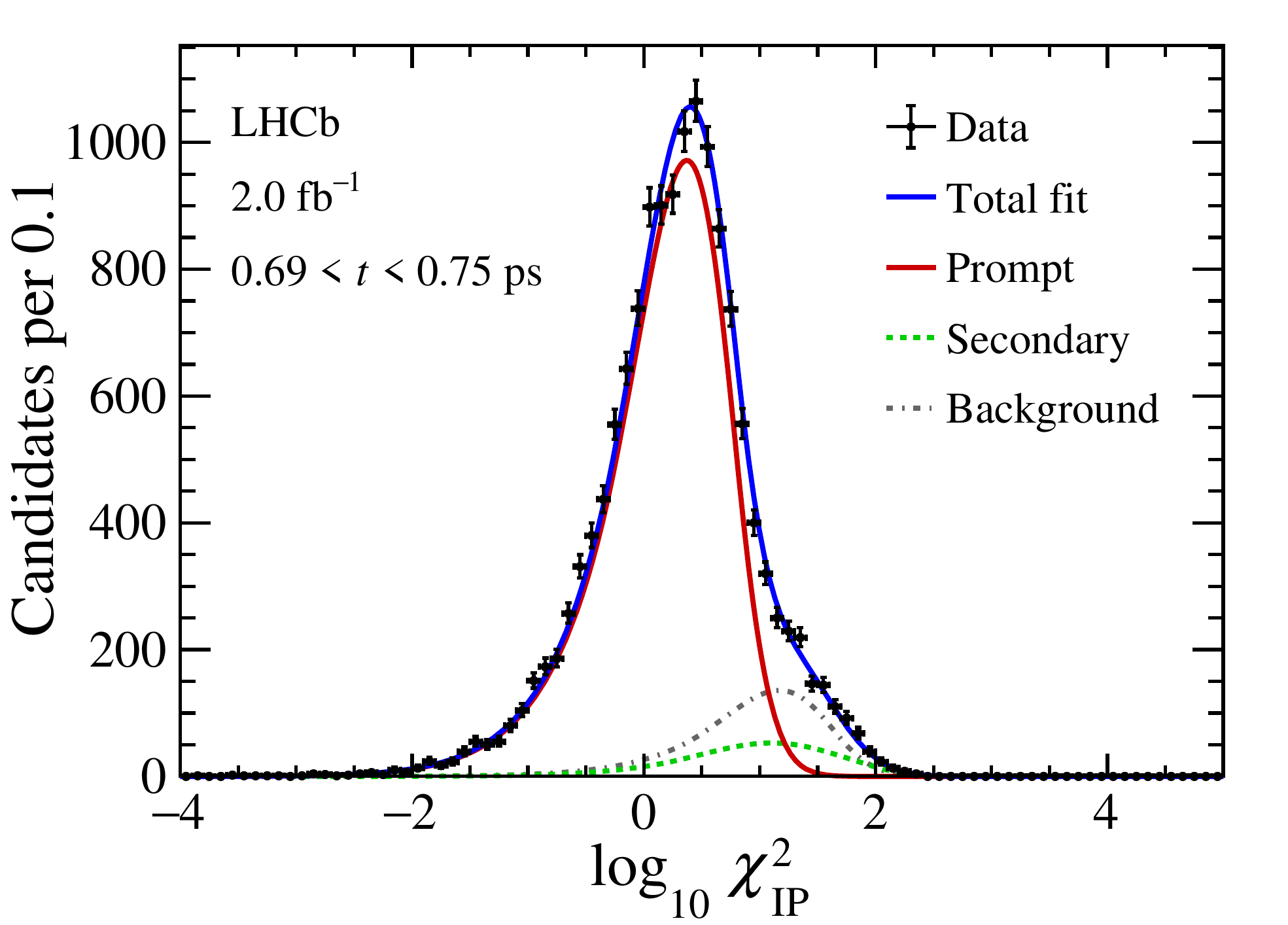}
  \caption{}
  \end{subfigure}
  \caption{(Color online)
  Distributions of 
  (a) invariant mass and
  (b) \logchisqip in the reduced mass region of $[2683,2707]\mevcc$
  for the \Omegac data sample,
  (c) invariant mass and
  (d) \logchisqip in the reduced mass region of $[2461,2481]\mevcc$
  for the \Xicz data sample,
  (e) invariant mass and
  (f) \logchisqip in the reduced mass region of $[1853, 1877]\mevcc$
  for the \Dz data sample,
  along with the fit results.
  The sample is collected in 2018\ in the decay-time interval of $[0.69,0.75]\ps$.
  The contributions of the signal, the secondary decays,
  and the combinatorial background are shown in red (solid), green (dashed), and gray (dash-dotted), respectively.
  }
  \label{fig:twod_fit}
\end{figure}

Example distributions of invariant mass and \logchisqip, in reduced mass regions around the peak, are presented in Fig.~\ref{fig:twod_fit}; the decay-time interval of $[0.69,0.75]\ps$ for data collected in 2018 is shown.
To obtain the \Omegac, \Xicz, and \Dz signal yields in each decay-time interval,
two-dimensional unbinned extended maximum likelihood fits
are performed to the invariant-mass and $\logchisqip$ distributions.
For each mode, the fits are performed simultaneously in all decay-time intervals and the three data-taking periods, 2016, 2017, and 2018.
The invariant-mass distribution of the signal candidates is described with the sum of a Gaussian function
and a double-sided Crystal Ball function~\cite{Skwarnicki:1986xj} with a shared mean.
The fit parameters are fixed to values obtained from simulation except for the mass peak
and the effective resolution,
which are obtained directly from data,
but shared among the different decay-time intervals.
The invariant-mass distribution of the combinatorial background contribution 
is described by a linear function with a slope left free to vary in the fit.
The $\logchisqip$ distributions of both the signal and background components
are described by a Bukin function~\cite{Bukin:2007}.
For signal components, parameters of the Bukin function 
are fixed to values obtained from simulation except for the peak position
that depends on the decay time.
Here, an offset parameter is added
to account for the disagreement between data and simulation.
The offset parameter is free to vary and shared between decay-time intervals 
and data-taking periods in the fit.
The parameters of the Bukin function of the combinatorial background contribution 
are fixed to values obtained from fits to the data samples 
in the sideband region as defined in the BDT training.
The two-dimensional model used for signal, secondary decays, and background components
is the product of the models 
for the invariant-mass and for the \logchisqip distributions.
Fit projections to the invariant-mass and \logchisqip distribution 
are shown in Fig~\ref{fig:twod_fit}.

\section{Decay time fit}
\begin{figure}[tb!]
  \centering
  \begin{subfigure}[t]{0.45\textwidth}
  \includegraphics[width=\linewidth]{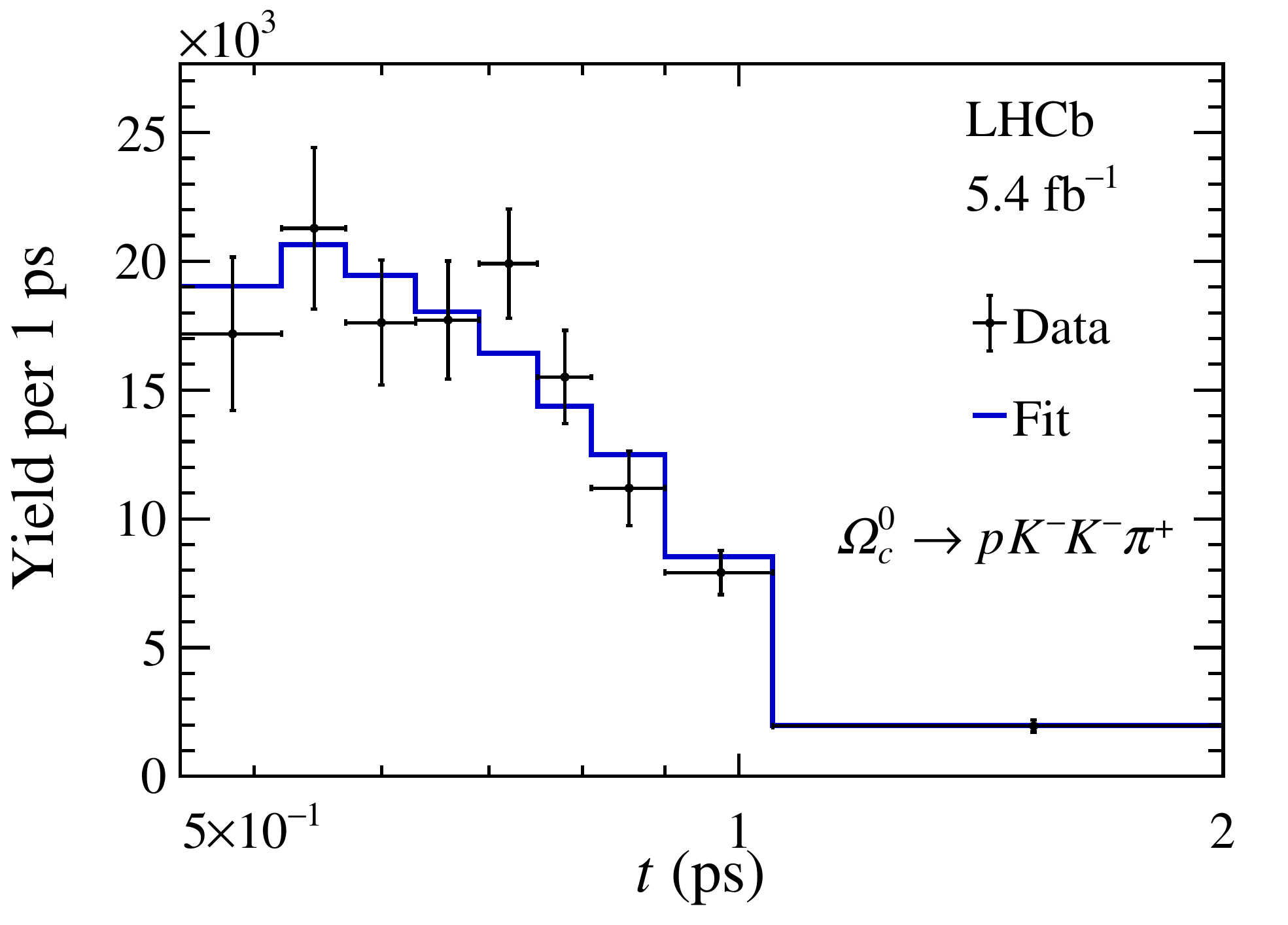}
  \caption{}
  \end{subfigure}
  \begin{subfigure}[t]{0.45\textwidth}
  \includegraphics[width=\linewidth]{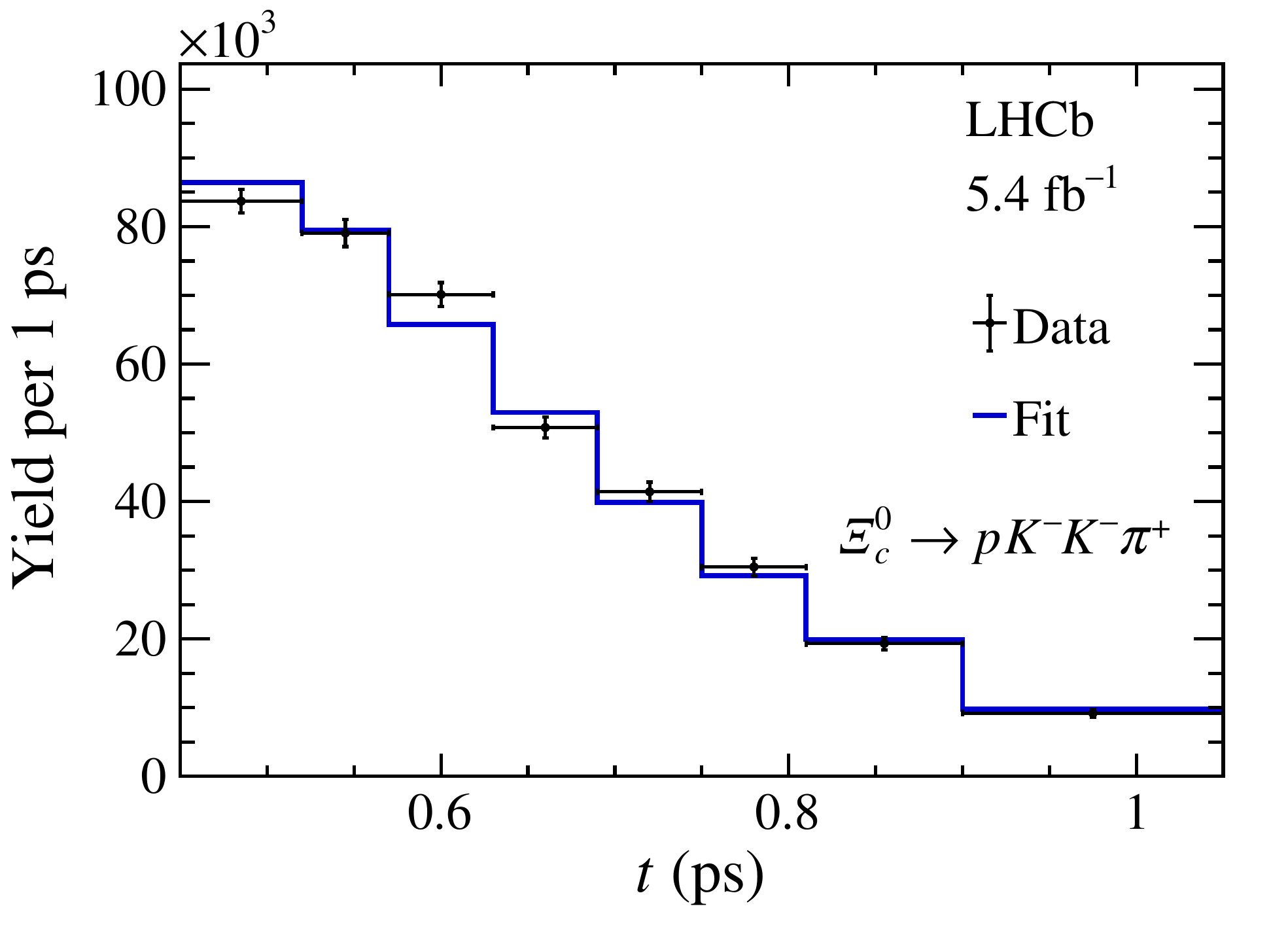}
  \caption{}
  \end{subfigure}
  \caption{(Color online)
  Decay-time distributions for
  the (a) \Omegac mode and the (b) \Xicz mode with the \chisq fit superimposed.
  The uncertainty on the data distribution is statistical only.}
  \label{fig:decay_time}
\end{figure}

The lifetimes of the \Omegac and \Xicz baryons are determined from 
a binned \chisq fit comparing the signal yields in data with those from the simulation, where the lifetime is known.
The latter is corrected using the control mode, as follows
\begin{equation}
\begin{aligned}
  \label{eq:decaytime_chisq}
  \chisq(\tau,\vec{C}) &= 
  \sum_{j}
  \sum_i \frac{\left(N^{\mathrm{sig}}_{i,j} - C_{j} 
               \times F_i(\tau) \times 
               R_{i,j} \right)^2}
    { \sigma^2_{N^{\mathrm{sig}}_{i,j}} + 
      C^2_{j} \times F_i^2(\tau) \times 
      \sigma^2_{R_{i,j}} },
      \qquad
      R_{i,j} &= \frac{N^{\mathrm{con}}_{i,j}}{M^{\mathrm{con}}_{i,j}} 
      \times M^{\mathrm{sig}}_{i,j},
\end{aligned}
\end{equation}
where: $N^{\mathrm{sig}}_{i,j}$ ($N^{\mathrm{con}}_{i,j}$)
is the signal yield in data for the signal (control) mode
in decay-time interval $i$ and
for the data-taking period $j$;
$M_{i,j}$ is the effective yield predicted from simulation;
$C_j$ is a normalisation factor to account for the difference in size between
the data and the simulated samples;
and $\sigma$ is the uncertainty of the relevant quantity.
The difference in lifetime between data and simulated samples is accounted for by
\begin{equation}
F_i(\tau) = \frac{\int_i \exp(-t/\tau) \mathrm{d}t}
                  {\int_i \exp(-t/\tau_{\mathrm{sim}}) \mathrm{d}t} \times
             \frac{\int_i \exp(-t/\tau^{\mathrm{con}}_{\mathrm{sim}}) \mathrm{d}t}
                  {\int_i \exp(-t/\tau^{\mathrm{con}}) \mathrm{d}t},
\end{equation}
where $\tau_{\mathrm{sim}} = 250\fs$ is the signal mode lifetime in simulation
and $\tau^{\mathrm{con}} = \tau^{\mathrm{con}}_{\mathrm{sim}}$ is
the known \Dz lifetime~\cite{PDG2020}, but is allowed to vary for estimating the systematic uncertainty.
The resulting lifetime is $\tau_{\Omegac} = \omegacval \pm \omegacstat \fs$
with $\chisqndf = 22/23$
and $\tau_{\Xicz} = \xiczval \pm \xiczstat \fs$
with $\chisqndf = 30/20$,
where the uncertainty is due to 
the limited size of the data and simulation samples.
The result of the \chisq fit to data is illustrated in Fig.~\ref{fig:decay_time}, which shows the 
signal yield $N^{\mathrm{sig}}$ for selected candidates as a function of decay time,
divided by the width of the corresponding decay-time interval,
where the fit results are superimposed.

Several cross-checks are performed to ensure the robustness of the results.
The \chisq fit is performed to data of the $\Dz\to\kkpipi$ control mode
for each data-taking period to validate the analysis procedure.
The obtained lifetimes are consistent between data-taking periods
and with the known \Dz lifetime~\cite{PDG2020}.
The data samples are split into sub-samples according to data-taking periods
and magnetic polarities of the LHCb dipole magnet,
and the lifetimes are measured for each sub-sample.
The resulting lifetimes are in good agreement with each other 
and with the default results.
The measurement is repeated with two alternative boundaries of decay-time intervals
and the obtained lifetimes are consistent with the default results within their statistical uncertainties.
To ensure that the result is independent of the input lifetime used in simulation,
the simulated signal decays are weighted to have alternative effective lifetimes 
within seven times the statistical uncertainty around the default lifetime.
The \chisq fit is then repeated.
The difference of the obtained lifetimes with regard to the default fit is negligible.

\section{Systematic uncertainties}
Sources of systematic uncertainty are investigated and
summarised in Table~\ref{tab:syst},
including those due to the fit model, the limited size of the calibration samples, 
differences between data and simulation, and the uncertainty due to the choice of the \Dz control mode. 
\begin{table}[b]
  \centering
  \caption{Systematic uncertainties for the \Omegac and \Xicz lifetimes.}
  \label{tab:syst}
  \begin{tabular}{l  cc}
    \toprule
    Sources & $\tau_\Omegac$ [fs] & $\tau_\Xicz$ [fs] \\
    \midrule
    Fit model                     & \phantom{0}2.2  & 1.0 \\
    Calibration sample size       & \phantom{0}0.1  & 0.1 \\
    Kinematic correction          & \phantom{0}3.4  & 0.4 \\
    Decay-time resolution         & \phantom{0}1.3  & 1.8 \\
    \chisqip scaling              & \phantom{0}1.1  & 0.5 \\
    Decay-length scale            & \phantom{0}0.1  & 0.1 \\
    \Dz--\Dzb mixing              & \phantom{0}0.8  & 0.6 \\
    \midrule
    Total systematic uncertainty  & \phantom{0}\omegacsys  & \xiczsys \\
    \midrule
    \Dz lifetime                  & \phantom{0}0.7  & 0.2 \\
    \midrule
    Statistical uncertainty       & \omegacstat & \xiczstat \\
    \bottomrule
  \end{tabular}
\end{table}
The systematic uncertainty due to the modelling of \logchisqip is 
studied with the \Dz control mode.
The following alternative models were tried and their impact on the signal yields studied. 
First, the effect due to fixed parameters in the Bukin function
is studied by removing these constraints one at a time
in the fit to the invariant-mass and $\logchisqip$ distributions.
Second, the uncertainty due to the choice of a single offset parameter 
for the peak positions of the Bukin functions across different decay-time intervals 
is studied by allowing independent offsets in each decay-time interval.
Third, an alternative model for the \logchisqip distribution of the combinatorial background
is obtained with the \sPlot technique~\cite{Pivk:2004ty}
using the invariant mass as the discriminating variable.
Half of the largest difference between the signal yields from the alternative model fits
is taken as the systematic uncertainty.
The obtained systematic uncertainties on the signal yields
are propagated to the measured lifetime using pseudoexperiments.
In each pseudoexperiment, the yields of the signal and control modes 
are varied according to a Gaussian distribution
whose mean is the value obtained with the default fit model and 
standard deviation the systematic uncertainty obtained with alternative models in the corresponding decay-time interval, and the lifetime is fit.
The standard deviation of the distribution for the fitted lifetime is taken as the 
systematic uncertainty.

The selection efficiency of the hardware trigger is estimated in data 
using \Lc candidates from semileptonic \Lb decays~\cite{MartinSanchez:1407893}.
The uncertainty due to the limited size of the calibration sample is estimated 
using pseudoexperiments, 
where the efficiency determined from the calibration sample
is varied according to its uncertainty.
The standard deviation of the distribution of the fitted lifetime is taken as the
systematic uncertainty.
The kinematic distributions of the simulated signal decays,
including the transverse momentum and rapidity of the charmed hadron 
and the transverse momentum of final-state tracks,
are weighted according to the distributions observed in data for each mode.
The impact of the limited size of the data samples,
which is more pronounced for the $\Omegac$ mode,
is studied with pseudoexperiments following the same procedure 
as described above.

Decay-time resolution in data is known to be different from simulation, although it cannot be accurately determined for the signal modes due to their limited yields in data.
Nonetheless, the impact of this difference on the measured charm-hadron lifetimes largely cancels due to taking the ratio with the \Dz control mode.
The residual effect is studied using pseudoexperiments and assigned as a systematic uncertainty.
For these pseudoexperiments the \Dz control mode is generated with both a 30\% larger and smaller decay-time resolution in the pseudo-data compared to pseudo-simulation, and the lifetime is fit.
The difference between the input lifetime and 
the mean value of the distribution of the fitted lifetimes is taken as the systematic uncertainty.

The \chisqip variables of the final-state tracks in simulation
are scaled to account for differences between data and simulation
for the data-taking periods of 2017 and 2018.
The scaling factor is obtained by comparing data distributions in the control mode.
The uncertainty on the scaling factor is determined to be $2\%$,
based on \chisq comparisons of data distributions with alternate scaling factors.
The difference between the default fitted lifetime and 
the lifetime determined with a scaling factor varied by $2\%$
is taken as the systematic uncertainty.

The measurement of the distance between the PV and the charmed-hadron decay vertex
depends on the relative longitudinal positions of the vertex locator modules of the \lhcb detector with respect to the beam axis.
The uncertainty on the positions of the modules is estimated using survey measurements and 
the track based alignment~\cite{HULSBERGEN2009471,VIRET2008157}, where the latter has the larger contribution.
Its uncertainty does not cancel in the decay-time ratio and is taken as a relative systematic uncertainty
of the measured lifetime.

The \Dz signal decays are reconstructed in a self-conjugate final state 
and \Dz--\Dzb mixing is not considered in the \chisq fit of the lifetime.
The impact of \Dz mixing is estimated using pseudoexperiments
in which the \Dz decay-time distribution is generated with mixing terms and the default \chisq fit is
performed to obtain the lifetime. The obtained difference between the input and resultant lifetime
is assigned as a systematic uncertainty.

The known value of $410.1\fs$~\cite{PDG2020} is assigned as \Dz lifetime
in the default decay-time fit.
The uncertainty on the \Dz lifetime, $1.5\fs$~\cite{PDG2020},
is propagated to the measured lifetime using pseudoexperiments. 
In each pseudoexperiment, the \Dz lifetime is varied according to its uncertainty.
The standard deviation of the distribution for the fitted lifetime is taken as the
systematic uncertainty.

\section{Conclusion}
In summary, a measurement of the lifetimes of the \Omegac and \Xicz baryons
is reported with  \Omegac and \Xicz baryons produced directly in proton--proton collisions 
at a centre-of-mass energy of $13\tev$,
corresponding to an integrated luminosity of $5.4\invfb$
collected by the \lhcb experiment.
The \Omegac lifetime is measured to be
\begin{equation*}
 \tau_{\Omegac} = \omegacval\pm\omegacstat\pm\omegacsys\pm\omegacext\fs,
\end{equation*}
and the \Xicz lifetime is measured to be
\begin{equation*}
\tau_{\Xicz} = \xiczval\pm\xiczstat\pm\xiczsys\pm\xiczext\fs,
\end{equation*}
where the first uncertainty is statistical, the second systematic,
and the third due to the uncertainty of the \Dz lifetime.
This result is consistent with the previous \lhcb measurements of 
the \Omegac and \Xicz lifetimes, obtained from semileptonic beauty-hadron decays~\cite{LHCb-PAPER-2018-028,LHCb-PAPER-2019-008},
and confirms the charmed-hadron lifetime hierarchy of \mbox{$\tau_{\Xicp}>\tau_{\Omegac}>\tau_{\Lc}>\tau_{\Xicz}$}.
The precision of the \Omegac lifetime is improved by a factor of two compared to that of the previous result~\cite{LHCb-PAPER-2018-028}.

This result is independent of previous \lhcb measurements~\cite{LHCb-PAPER-2018-028,LHCb-PAPER-2019-008}
due to the choice of independent data sample and analysis technique.
Combining this measurement with previous \lhcb measurements~\cite{LHCb-PAPER-2018-028,LHCb-PAPER-2019-008},
given that both the statistical uncertainties 
and the dominant systematic uncertainties are uncorrelated,
results in the weighted average lifetimes of
\begin{equation*}
  \begin{aligned}
    \tau_{\Omegac} &= \combomegacval \pm \combomegacerr \fs,\\
    \tau_{\Xicz}   &= \combxiczval \pm \combxiczerr \fs.
  \end{aligned}
\end{equation*}
The uncertainty includes both the statistical and systematic uncertainties.

\section*{Acknowledgements}
%
%
\noindent We express our gratitude to our colleagues in the CERN
accelerator departments for the excellent performance of the LHC. We
thank the technical and administrative staff at the LHCb
institutes.
We acknowledge support from CERN and from the national agencies:
CAPES, CNPq, FAPERJ and FINEP (Brazil); 
MOST and NSFC (China); 
CNRS/IN2P3 (France); 
BMBF, DFG and MPG (Germany); 
INFN (Italy); 
NWO (Netherlands); 
MNiSW and NCN (Poland); 
MEN/IFA (Romania); 
MSHE (Russia); 
MICINN (Spain); 
SNSF and SER (Switzerland); 
NASU (Ukraine); 
STFC (United Kingdom); 
DOE NP and NSF (USA).
We acknowledge the computing resources that are provided by CERN, IN2P3
(France), KIT and DESY (Germany), INFN (Italy), SURF (Netherlands),
PIC (Spain), GridPP (United Kingdom), RRCKI and Yandex
LLC (Russia), CSCS (Switzerland), IFIN-HH (Romania), CBPF (Brazil),
PL-GRID (Poland) and OSC (USA).
We are indebted to the communities behind the multiple open-source
software packages on which we depend.
Individual groups or members have received support from
ARC and ARDC (Australia);
AvH Foundation (Germany);
EPLANET, Marie Sk\l{}odowska-Curie Actions and ERC (European Union);
A*MIDEX, ANR, Labex P2IO and OCEVU, and R\'{e}gion Auvergne-Rh\^{o}ne-Alpes (France);
Key Research Program of Frontier Sciences of CAS, CAS PIFI, CAS CCEPP, 
Fundamental Research Funds for the Central Universities, 
and Sci. \& Tech. Program of Guangzhou (China);
RFBR, RSF and Yandex LLC (Russia);
GVA, XuntaGal and GENCAT (Spain);
the Royal Society
and the Leverhulme Trust (United Kingdom).

\addcontentsline{toc}{section}{References}
\bibliographystyle{LHCb}
\bibliography{main,standard,LHCb-PAPER,LHCb-CONF,LHCb-DP,LHCb-TDR}

\ifx\mcitethebibliography\mciteundefinedmacro
\PackageError{LHCb.bst}{mciteplus.sty has not been loaded}
{This bibstyle requires the use of the mciteplus package.}\fi
\providecommand{\href}[2]{#2}
\begin{mcitethebibliography}{10}
\mciteSetBstSublistMode{n}
\mciteSetBstMaxWidthForm{subitem}{\alph{mcitesubitemcount})}
\mciteSetBstSublistLabelBeginEnd{\mcitemaxwidthsubitemform\space}
{\relax}{\relax}

\bibitem{LHCb-PAPER-2018-028}
LHCb collaboration, R.~Aaij {\em et~al.},
  \ifthenelse{\boolean{articletitles}}{\emph{{Measurement of the
  $\Omegares_c^0$ lifetime}},
  }{}\href{https://doi.org/10.1103/PhysRevLett.121.092003}{Phys.\ Rev.\ Lett.\
  \textbf{121} (2018) 092003},
  \href{http://arxiv.org/abs/1807.02024}{{\normalfont\ttfamily
  arXiv:1807.02024}}\relax
\mciteBstWouldAddEndPuncttrue
\mciteSetBstMidEndSepPunct{\mcitedefaultmidpunct}
{\mcitedefaultendpunct}{\mcitedefaultseppunct}\relax
\EndOfBibitem
\bibitem{LHCb-PAPER-2019-008}
LHCb collaboration, R.~Aaij {\em et~al.},
  \ifthenelse{\boolean{articletitles}}{\emph{{Precision measurement of the \Lc,
  \Xicp, and \Xicz baryon lifetimes}},
  }{}\href{https://doi.org/10.1103/PhysRevD.100.032001}{Phys.\ Rev.\
  \textbf{D100} (2019) 032001},
  \href{http://arxiv.org/abs/1906.08350}{{\normalfont\ttfamily
  arXiv:1906.08350}}\relax
\mciteBstWouldAddEndPuncttrue
\mciteSetBstMidEndSepPunct{\mcitedefaultmidpunct}
{\mcitedefaultendpunct}{\mcitedefaultseppunct}\relax
\EndOfBibitem
\bibitem{PDG2018}
Particle Data Group, M.~Tanabashi {\em et~al.},
  \ifthenelse{\boolean{articletitles}}{\emph{{\href{http://pdg.lbl.gov/}{Review
  of particle physics}}},
  }{}\href{https://doi.org/10.1103/PhysRevD.98.030001}{Phys.\ Rev.\
  \textbf{D98} (2018) 030001}\relax
\mciteBstWouldAddEndPuncttrue
\mciteSetBstMidEndSepPunct{\mcitedefaultmidpunct}
{\mcitedefaultendpunct}{\mcitedefaultseppunct}\relax
\EndOfBibitem
\bibitem{Khoze:1983yp}
V.~A. Khoze and M.~A. Shifman,
  \ifthenelse{\boolean{articletitles}}{\emph{{Heavy quarks}},
  }{}\href{https://doi.org/10.1070/PU1983v026n05ABEH004398}{Sov.\ Phys.\ Usp.\
  \textbf{26} (1983) 387}\relax
\mciteBstWouldAddEndPuncttrue
\mciteSetBstMidEndSepPunct{\mcitedefaultmidpunct}
{\mcitedefaultendpunct}{\mcitedefaultseppunct}\relax
\EndOfBibitem
\bibitem{Bigi:1991ir}
I.~I.~Y. Bigi and N.~G. Uraltsev,
  \ifthenelse{\boolean{articletitles}}{\emph{{Gluonic enhancements in
  non-spectator beauty decays: an inclusive mirage though an exclusive
  possibility}}, }{}\href{https://doi.org/10.1016/0370-2693(92)90066-D}{Phys.\
  Lett.\  \textbf{B280} (1992) 271}\relax
\mciteBstWouldAddEndPuncttrue
\mciteSetBstMidEndSepPunct{\mcitedefaultmidpunct}
{\mcitedefaultendpunct}{\mcitedefaultseppunct}\relax
\EndOfBibitem
\bibitem{Bigi:1992su}
I.~I.~Y. Bigi, N.~G. Uraltsev, and A.~I. Vainshtein,
  \ifthenelse{\boolean{articletitles}}{\emph{{Nonperturbative corrections to
  inclusive beauty and charm decays: QCD versus phenomenological models}},
  }{}\href{https://doi.org/10.1016/0370-2693(92)90908-M}{Phys.\ Lett.\
  \textbf{B293} (1992) 430},
  \href{http://arxiv.org/abs/hep-ph/9207214}{{\normalfont\ttfamily
  arXiv:hep-ph/9207214}}, [Erratum: Phys. Lett. B297, (1992) 477--477]\relax
\mciteBstWouldAddEndPuncttrue
\mciteSetBstMidEndSepPunct{\mcitedefaultmidpunct}
{\mcitedefaultendpunct}{\mcitedefaultseppunct}\relax
\EndOfBibitem
\bibitem{Blok:1992hw}
B.~Blok and M.~A. Shifman, \ifthenelse{\boolean{articletitles}}{\emph{{The rule
  of discarding $1/N_c$ in inclusive weak decays. (I)}},
  }{}\href{https://doi.org/10.1016/0550-3213(93)90504-I}{Nucl.\ Phys.\
  \textbf{B399} (1993) 441},
  \href{http://arxiv.org/abs/hep-ph/9207236}{{\normalfont\ttfamily
  arXiv:hep-ph/9207236}}\relax
\mciteBstWouldAddEndPuncttrue
\mciteSetBstMidEndSepPunct{\mcitedefaultmidpunct}
{\mcitedefaultendpunct}{\mcitedefaultseppunct}\relax
\EndOfBibitem
\bibitem{Blok:1992he}
B.~Blok and M.~A. Shifman, \ifthenelse{\boolean{articletitles}}{\emph{{The rule
  of discarding $1/N_c$ in inclusive weak decays. (II)}},
  }{}\href{https://doi.org/10.1016/0550-3213(93)90505-J}{Nucl.\ Phys.\
  \textbf{B399} (1993) 459},
  \href{http://arxiv.org/abs/hep-ph/9209289}{{\normalfont\ttfamily
  arXiv:hep-ph/9209289}}\relax
\mciteBstWouldAddEndPuncttrue
\mciteSetBstMidEndSepPunct{\mcitedefaultmidpunct}
{\mcitedefaultendpunct}{\mcitedefaultseppunct}\relax
\EndOfBibitem
\bibitem{Neubert:1997gu}
M.~Neubert, \ifthenelse{\boolean{articletitles}}{\emph{{B decays and the heavy
  quark expansion}}, }{}\href{https://doi.org/10.1142/9789812812667_0003}{Adv.\
  Ser.\ Direct.\ High Energy Phys.\  \textbf{15} (1998) 239},
  \href{http://arxiv.org/abs/hep-ph/9702375}{{\normalfont\ttfamily
  arXiv:hep-ph/9702375}}\relax
\mciteBstWouldAddEndPuncttrue
\mciteSetBstMidEndSepPunct{\mcitedefaultmidpunct}
{\mcitedefaultendpunct}{\mcitedefaultseppunct}\relax
\EndOfBibitem
\bibitem{Uraltsev:1998bk}
N.~Uraltsev, \ifthenelse{\boolean{articletitles}}{\emph{{Heavy quark expansion
  in beauty and its decays}},
  }{}\href{https://doi.org/10.3254/978-1-61499-222-6-329}{Proc.\ Int.\ Sch.\
  Phys.\ Fermi \textbf{137} (1998) 329},
  \href{http://arxiv.org/abs/hep-ph/9804275}{{\normalfont\ttfamily
  arXiv:hep-ph/9804275}}\relax
\mciteBstWouldAddEndPuncttrue
\mciteSetBstMidEndSepPunct{\mcitedefaultmidpunct}
{\mcitedefaultendpunct}{\mcitedefaultseppunct}\relax
\EndOfBibitem
\bibitem{Bigi:1995jr}
I.~I.~Y. Bigi, \ifthenelse{\boolean{articletitles}}{\emph{{The QCD perspective
  on lifetimes of heavy flavor hadrons}},
  }{}\href{http://arxiv.org/abs/hep-ph/9508408}{{\normalfont\ttfamily
  arXiv:hep-ph/9508408}}\relax
\mciteBstWouldAddEndPuncttrue
\mciteSetBstMidEndSepPunct{\mcitedefaultmidpunct}
{\mcitedefaultendpunct}{\mcitedefaultseppunct}\relax
\EndOfBibitem
\bibitem{Kirk:2017juj}
M.~Kirk, A.~Lenz, and T.~Rauh,
  \ifthenelse{\boolean{articletitles}}{\emph{{Dimension-six matrix elements for
  meson mixing and lifetimes from sum rules}},
  }{}\href{https://doi.org/10.1007/JHEP12(2017)068}{JHEP \textbf{12} (2017)
  068}, \href{http://arxiv.org/abs/1711.02100}{{\normalfont\ttfamily
  arXiv:1711.02100}}, [Erratum: JHEP 06, 162 (2020)]\relax
\mciteBstWouldAddEndPuncttrue
\mciteSetBstMidEndSepPunct{\mcitedefaultmidpunct}
{\mcitedefaultendpunct}{\mcitedefaultseppunct}\relax
\EndOfBibitem
\bibitem{Cheng:2015iom}
H.-Y. Cheng, \ifthenelse{\boolean{articletitles}}{\emph{{Charmed baryons circa
  2015}}, }{}\href{https://doi.org/10.1007/s11467-015-0483-z}{Front.\ Phys.\
  (Beijing) \textbf{10} (2015) 101406}\relax
\mciteBstWouldAddEndPuncttrue
\mciteSetBstMidEndSepPunct{\mcitedefaultmidpunct}
{\mcitedefaultendpunct}{\mcitedefaultseppunct}\relax
\EndOfBibitem
\bibitem{Lenz:2013aua}
A.~Lenz and T.~Rauh, \ifthenelse{\boolean{articletitles}}{\emph{{$D$-meson
  lifetimes within the heavy quark expansion}},
  }{}\href{https://doi.org/10.1103/PhysRevD.88.034004}{Phys.\ Rev.\
  \textbf{D88} (2013) 034004},
  \href{http://arxiv.org/abs/1305.3588}{{\normalfont\ttfamily
  arXiv:1305.3588}}\relax
\mciteBstWouldAddEndPuncttrue
\mciteSetBstMidEndSepPunct{\mcitedefaultmidpunct}
{\mcitedefaultendpunct}{\mcitedefaultseppunct}\relax
\EndOfBibitem
\bibitem{Bianco:2003vb}
S.~Bianco, F.~L. Fabbri, D.~Benson, and I.~I.~Y. Bigi,
  \ifthenelse{\boolean{articletitles}}{\emph{{A Cicerone for the physics of
  charm}}, }{}\href{https://doi.org/10.1393/ncr/i2003-10003-1}{Riv.\ Nuovo
  Cim.\  \textbf{26} (2003) 1},
  \href{http://arxiv.org/abs/hep-ex/0309021}{{\normalfont\ttfamily
  arXiv:hep-ex/0309021}}\relax
\mciteBstWouldAddEndPuncttrue
\mciteSetBstMidEndSepPunct{\mcitedefaultmidpunct}
{\mcitedefaultendpunct}{\mcitedefaultseppunct}\relax
\EndOfBibitem
\bibitem{Bellini:1996ra}
G.~Bellini, I.~I.~Y. Bigi, and P.~J. Dornan,
  \ifthenelse{\boolean{articletitles}}{\emph{{Lifetimes of charm and beauty
  hadrons}}, }{}\href{https://doi.org/10.1016/S0370-1573(97)00005-7}{Phys.\
  Rept.\  \textbf{289} (1997) 1}\relax
\mciteBstWouldAddEndPuncttrue
\mciteSetBstMidEndSepPunct{\mcitedefaultmidpunct}
{\mcitedefaultendpunct}{\mcitedefaultseppunct}\relax
\EndOfBibitem
\bibitem{Blok:1991st}
B.~Blok and M.~A. Shifman,
  \ifthenelse{\boolean{articletitles}}{\emph{{Lifetimes of charmed hadrons
  revisited. Facts and fancy}}, }{} in {\em {3rd Workshop on the Tau-Charm
  Factory}}, 1991,
  \href{http://arxiv.org/abs/hep-ph/9311331}{{\normalfont\ttfamily
  arXiv:hep-ph/9311331}}\relax
\mciteBstWouldAddEndPuncttrue
\mciteSetBstMidEndSepPunct{\mcitedefaultmidpunct}
{\mcitedefaultendpunct}{\mcitedefaultseppunct}\relax
\EndOfBibitem
\bibitem{Cheng:1997xba}
H.-Y. Cheng, \ifthenelse{\boolean{articletitles}}{\emph{{A Phenomenological
  analysis of heavy hadron lifetimes}},
  }{}\href{https://doi.org/10.1103/PhysRevD.56.2783}{Phys.\ Rev.\  \textbf{D56}
  (1997) 2783},
  \href{http://arxiv.org/abs/hep-ph/9704260}{{\normalfont\ttfamily
  arXiv:hep-ph/9704260}}\relax
\mciteBstWouldAddEndPuncttrue
\mciteSetBstMidEndSepPunct{\mcitedefaultmidpunct}
{\mcitedefaultendpunct}{\mcitedefaultseppunct}\relax
\EndOfBibitem
\bibitem{Cheng:2018rkz}
H.-Y. Cheng, \ifthenelse{\boolean{articletitles}}{\emph{{Phenomenological Study
  of Heavy Hadron Lifetimes}},
  }{}\href{https://doi.org/10.1007/JHEP11(2018)014}{JHEP \textbf{11} (2018)
  014}, \href{http://arxiv.org/abs/1807.00916}{{\normalfont\ttfamily
  arXiv:1807.00916}}\relax
\mciteBstWouldAddEndPuncttrue
\mciteSetBstMidEndSepPunct{\mcitedefaultmidpunct}
{\mcitedefaultendpunct}{\mcitedefaultseppunct}\relax
\EndOfBibitem
\bibitem{cheng2021charmed}
H.-Y. Cheng, \ifthenelse{\boolean{articletitles}}{\emph{Charmed baryon physics
  circa 2021}, }{}\href{http://arxiv.org/abs/2109.01216}{{\normalfont\ttfamily
  arXiv:2109.01216}}\relax
\mciteBstWouldAddEndPuncttrue
\mciteSetBstMidEndSepPunct{\mcitedefaultmidpunct}
{\mcitedefaultendpunct}{\mcitedefaultseppunct}\relax
\EndOfBibitem
\bibitem{LHCb-DP-2008-001}
LHCb collaboration, A.~A. Alves~Jr.\ {\em et~al.},
  \ifthenelse{\boolean{articletitles}}{\emph{{The \lhcb detector at the LHC}},
  }{}\href{https://doi.org/10.1088/1748-0221/3/08/S08005}{JINST \textbf{3}
  (2008) S08005}\relax
\mciteBstWouldAddEndPuncttrue
\mciteSetBstMidEndSepPunct{\mcitedefaultmidpunct}
{\mcitedefaultendpunct}{\mcitedefaultseppunct}\relax
\EndOfBibitem
\bibitem{LHCb-DP-2014-002}
LHCb collaboration, R.~Aaij {\em et~al.},
  \ifthenelse{\boolean{articletitles}}{\emph{{LHCb detector performance}},
  }{}\href{https://doi.org/10.1142/S0217751X15300227}{Int.\ J.\ Mod.\ Phys.\
  \textbf{A30} (2015) 1530022},
  \href{http://arxiv.org/abs/1412.6352}{{\normalfont\ttfamily
  arXiv:1412.6352}}\relax
\mciteBstWouldAddEndPuncttrue
\mciteSetBstMidEndSepPunct{\mcitedefaultmidpunct}
{\mcitedefaultendpunct}{\mcitedefaultseppunct}\relax
\EndOfBibitem
\bibitem{LHCb-DP-2012-004}
R.~Aaij {\em et~al.}, \ifthenelse{\boolean{articletitles}}{\emph{{The \lhcb
  trigger and its performance in 2011}},
  }{}\href{https://doi.org/10.1088/1748-0221/8/04/P04022}{JINST \textbf{8}
  (2013) P04022}, \href{http://arxiv.org/abs/1211.3055}{{\normalfont\ttfamily
  arXiv:1211.3055}}\relax
\mciteBstWouldAddEndPuncttrue
\mciteSetBstMidEndSepPunct{\mcitedefaultmidpunct}
{\mcitedefaultendpunct}{\mcitedefaultseppunct}\relax
\EndOfBibitem
\bibitem{LHCb-DP-2016-001}
R.~Aaij {\em et~al.}, \ifthenelse{\boolean{articletitles}}{\emph{{Tesla: an
  application for real-time data analysis in High Energy Physics}},
  }{}\href{https://doi.org/10.1016/j.cpc.2016.07.022}{Comput.\ Phys.\ Commun.\
  \textbf{208} (2016) 35},
  \href{http://arxiv.org/abs/1604.05596}{{\normalfont\ttfamily
  arXiv:1604.05596}}\relax
\mciteBstWouldAddEndPuncttrue
\mciteSetBstMidEndSepPunct{\mcitedefaultmidpunct}
{\mcitedefaultendpunct}{\mcitedefaultseppunct}\relax
\EndOfBibitem
\bibitem{Sjostrand:2006za}
T.~Sj\"{o}strand, S.~Mrenna, and P.~Skands,
  \ifthenelse{\boolean{articletitles}}{\emph{{PYTHIA 6.4 physics and manual}},
  }{}\href{https://doi.org/10.1088/1126-6708/2006/05/026}{JHEP \textbf{05}
  (2006) 026}, \href{http://arxiv.org/abs/hep-ph/0603175}{{\normalfont\ttfamily
  arXiv:hep-ph/0603175}}\relax
\mciteBstWouldAddEndPuncttrue
\mciteSetBstMidEndSepPunct{\mcitedefaultmidpunct}
{\mcitedefaultendpunct}{\mcitedefaultseppunct}\relax
\EndOfBibitem
\bibitem{Sjostrand:2007gs}
T.~Sj\"{o}strand, S.~Mrenna, and P.~Skands,
  \ifthenelse{\boolean{articletitles}}{\emph{{A brief introduction to PYTHIA
  8.1}}, }{}\href{https://doi.org/10.1016/j.cpc.2008.01.036}{Comput.\ Phys.\
  Commun.\  \textbf{178} (2008) 852},
  \href{http://arxiv.org/abs/0710.3820}{{\normalfont\ttfamily
  arXiv:0710.3820}}\relax
\mciteBstWouldAddEndPuncttrue
\mciteSetBstMidEndSepPunct{\mcitedefaultmidpunct}
{\mcitedefaultendpunct}{\mcitedefaultseppunct}\relax
\EndOfBibitem
\bibitem{LHCb-PROC-2010-056}
I.~Belyaev {\em et~al.}, \ifthenelse{\boolean{articletitles}}{\emph{{Handling
  of the generation of primary events in Gauss, the LHCb simulation
  framework}}, }{}\href{https://doi.org/10.1088/1742-6596/331/3/032047}{J.\
  Phys.\ Conf.\ Ser.\  \textbf{331} (2011) 032047}\relax
\mciteBstWouldAddEndPuncttrue
\mciteSetBstMidEndSepPunct{\mcitedefaultmidpunct}
{\mcitedefaultendpunct}{\mcitedefaultseppunct}\relax
\EndOfBibitem
\bibitem{Lange:2001uf}
D.~J. Lange, \ifthenelse{\boolean{articletitles}}{\emph{{The EvtGen particle
  decay simulation package}},
  }{}\href{https://doi.org/10.1016/S0168-9002(01)00089-4}{Nucl.\ Instrum.\
  Meth.\  \textbf{A462} (2001) 152}\relax
\mciteBstWouldAddEndPuncttrue
\mciteSetBstMidEndSepPunct{\mcitedefaultmidpunct}
{\mcitedefaultendpunct}{\mcitedefaultseppunct}\relax
\EndOfBibitem
\bibitem{davidson2015photos}
N.~Davidson, T.~Przedzinski, and Z.~Was,
  \ifthenelse{\boolean{articletitles}}{\emph{{PHOTOS interface in C++:
  Technical and physics documentation}},
  }{}\href{https://doi.org/https://doi.org/10.1016/j.cpc.2015.09.013}{Comp.\
  Phys.\ Comm.\  \textbf{199} (2016) 86},
  \href{http://arxiv.org/abs/1011.0937}{{\normalfont\ttfamily
  arXiv:1011.0937}}\relax
\mciteBstWouldAddEndPuncttrue
\mciteSetBstMidEndSepPunct{\mcitedefaultmidpunct}
{\mcitedefaultendpunct}{\mcitedefaultseppunct}\relax
\EndOfBibitem
\bibitem{Allison:2006ve}
Geant4 collaboration, J.~Allison {\em et~al.},
  \ifthenelse{\boolean{articletitles}}{\emph{{Geant4 developments and
  applications}}, }{}\href{https://doi.org/10.1109/TNS.2006.869826}{IEEE
  Trans.\ Nucl.\ Sci.\  \textbf{53} (2006) 270}\relax
\mciteBstWouldAddEndPuncttrue
\mciteSetBstMidEndSepPunct{\mcitedefaultmidpunct}
{\mcitedefaultendpunct}{\mcitedefaultseppunct}\relax
\EndOfBibitem
\bibitem{Agostinelli:2002hh}
Geant4 collaboration, S.~Agostinelli {\em et~al.},
  \ifthenelse{\boolean{articletitles}}{\emph{{Geant4: A simulation toolkit}},
  }{}\href{https://doi.org/10.1016/S0168-9002(03)01368-8}{Nucl.\ Instrum.\
  Meth.\  \textbf{A506} (2003) 250}\relax
\mciteBstWouldAddEndPuncttrue
\mciteSetBstMidEndSepPunct{\mcitedefaultmidpunct}
{\mcitedefaultendpunct}{\mcitedefaultseppunct}\relax
\EndOfBibitem
\bibitem{LHCb-PROC-2011-006}
M.~Clemencic {\em et~al.}, \ifthenelse{\boolean{articletitles}}{\emph{{The
  \lhcb simulation application, Gauss: Design, evolution and experience}},
  }{}\href{https://doi.org/10.1088/1742-6596/331/3/032023}{J.\ Phys.\ Conf.\
  Ser.\  \textbf{331} (2011) 032023}\relax
\mciteBstWouldAddEndPuncttrue
\mciteSetBstMidEndSepPunct{\mcitedefaultmidpunct}
{\mcitedefaultendpunct}{\mcitedefaultseppunct}\relax
\EndOfBibitem
\bibitem{LHCb-DP-2018-004}
D.~M{\"u}ller, M.~Clemencic, G.~Corti, and M.~Gersabeck,
  \ifthenelse{\boolean{articletitles}}{\emph{{ReDecay: A novel approach to
  speed up the simulation at LHCb}},
  }{}\href{https://doi.org/10.1140/epjc/s10052-018-6469-6}{Eur.\ Phys.\ J.\
  \textbf{C78} (2018) 1009},
  \href{http://arxiv.org/abs/1810.10362}{{\normalfont\ttfamily
  arXiv:1810.10362}}\relax
\mciteBstWouldAddEndPuncttrue
\mciteSetBstMidEndSepPunct{\mcitedefaultmidpunct}
{\mcitedefaultendpunct}{\mcitedefaultseppunct}\relax
\EndOfBibitem
\bibitem{Breiman}
L.~Breiman, J.~H. Friedman, R.~A. Olshen, and C.~J. Stone, {\em Classification
  and regression trees}, Wadsworth international group, Belmont, California,
  USA, 1984\relax
\mciteBstWouldAddEndPuncttrue
\mciteSetBstMidEndSepPunct{\mcitedefaultmidpunct}
{\mcitedefaultendpunct}{\mcitedefaultseppunct}\relax
\EndOfBibitem
\bibitem{AdaBoost}
Y.~Freund and R.~E. Schapire, \ifthenelse{\boolean{articletitles}}{\emph{A
  decision-theoretic generalization of on-line learning and an application to
  boosting}, }{}\href{https://doi.org/10.1006/jcss.1997.1504}{J.\ Comput.\
  Syst.\ Sci.\  \textbf{55} (1997) 119}\relax
\mciteBstWouldAddEndPuncttrue
\mciteSetBstMidEndSepPunct{\mcitedefaultmidpunct}
{\mcitedefaultendpunct}{\mcitedefaultseppunct}\relax
\EndOfBibitem
\bibitem{Hocker:2007ht}
H.~Voss, A.~Hoecker, J.~Stelzer, and F.~Tegenfeldt,
  \ifthenelse{\boolean{articletitles}}{\emph{{TMVA - Toolkit for Multivariate
  Data Analysis with ROOT}}, }{}\href{https://doi.org/10.22323/1.050.0040}{PoS
  \textbf{ACAT} (2007) 040}\relax
\mciteBstWouldAddEndPuncttrue
\mciteSetBstMidEndSepPunct{\mcitedefaultmidpunct}
{\mcitedefaultendpunct}{\mcitedefaultseppunct}\relax
\EndOfBibitem
\bibitem{TMVA4}
A.~Hoecker {\em et~al.}, \ifthenelse{\boolean{articletitles}}{\emph{{TMVA 4 ---
  Toolkit for Multivariate Data Analysis with ROOT, Users Guide}},
  }{}\href{http://arxiv.org/abs/physics/0703039}{{\normalfont\ttfamily
  arXiv:physics/0703039}}\relax
\mciteBstWouldAddEndPuncttrue
\mciteSetBstMidEndSepPunct{\mcitedefaultmidpunct}
{\mcitedefaultendpunct}{\mcitedefaultseppunct}\relax
\EndOfBibitem
\bibitem{PDG2020}
Particle Data Group, P.~A. Zyla {\em et~al.},
  \ifthenelse{\boolean{articletitles}}{\emph{{\href{http://pdg.lbl.gov/}{Review
  of particle physics}}}, }{}\href{https://doi.org/10.1093/ptep/ptaa104}{Prog.\
  Theor.\ Exp.\ Phys.\  \textbf{2020} (2020) 083C01}\relax
\mciteBstWouldAddEndPuncttrue
\mciteSetBstMidEndSepPunct{\mcitedefaultmidpunct}
{\mcitedefaultendpunct}{\mcitedefaultseppunct}\relax
\EndOfBibitem
\bibitem{matrix_net}
A.~Gulin, I.~Kuralenok, and D.~Pavlov, in {\em Winning the transfer learning
  track of Yahoo!'s learning to rank challenge with YetiRank}, 2011\relax
\mciteBstWouldAddEndPuncttrue
\mciteSetBstMidEndSepPunct{\mcitedefaultmidpunct}
{\mcitedefaultendpunct}{\mcitedefaultseppunct}\relax
\EndOfBibitem
\bibitem{LHCb-DP-2019-001}
R.~Aaij {\em et~al.}, \ifthenelse{\boolean{articletitles}}{\emph{{Performance
  of the LHCb trigger and full real-time reconstruction in Run 2 of the LHC}},
  }{}\href{https://doi.org/10.1088/1748-0221/14/04/P04013}{JINST \textbf{14}
  (2019) P04013}, \href{http://arxiv.org/abs/1812.10790}{{\normalfont\ttfamily
  arXiv:1812.10790}}\relax
\mciteBstWouldAddEndPuncttrue
\mciteSetBstMidEndSepPunct{\mcitedefaultmidpunct}
{\mcitedefaultendpunct}{\mcitedefaultseppunct}\relax
\EndOfBibitem
\bibitem{Skwarnicki:1986xj}
T.~Skwarnicki, {\em {A study of the radiative cascade transitions between the
  Upsilon-prime and Upsilon resonances}}, PhD thesis, Institute of Nuclear
  Physics, Krakow, 1986,
  {\href{http://inspirehep.net/record/230779/}{DESY-F31-86-02}}\relax
\mciteBstWouldAddEndPuncttrue
\mciteSetBstMidEndSepPunct{\mcitedefaultmidpunct}
{\mcitedefaultendpunct}{\mcitedefaultseppunct}\relax
\EndOfBibitem
\bibitem{Bukin:2007}
A.~D. Bukin, \ifthenelse{\boolean{articletitles}}{\emph{{Fitting function for
  asymmetric peaks}},
  }{}\href{http://arxiv.org/abs/0711.4449}{{\normalfont\ttfamily
  arXiv:0711.4449}}\relax
\mciteBstWouldAddEndPuncttrue
\mciteSetBstMidEndSepPunct{\mcitedefaultmidpunct}
{\mcitedefaultendpunct}{\mcitedefaultseppunct}\relax
\EndOfBibitem
\bibitem{Pivk:2004ty}
M.~Pivk and F.~R. Le~Diberder,
  \ifthenelse{\boolean{articletitles}}{\emph{{sPlot: A statistical tool to
  unfold data distributions}},
  }{}\href{https://doi.org/10.1016/j.nima.2005.08.106}{Nucl.\ Instrum.\ Meth.\
  \textbf{A555} (2005) 356},
  \href{http://arxiv.org/abs/physics/0402083}{{\normalfont\ttfamily
  arXiv:physics/0402083}}\relax
\mciteBstWouldAddEndPuncttrue
\mciteSetBstMidEndSepPunct{\mcitedefaultmidpunct}
{\mcitedefaultendpunct}{\mcitedefaultseppunct}\relax
\EndOfBibitem
\bibitem{MartinSanchez:1407893}
A.~Martin~Sanchez, P.~Robbe, and M.-H. Schune,
  \ifthenelse{\boolean{articletitles}}{\emph{{Performances of the LHCb L0
  Calorimeter Trigger}}}{},
  \href{https://cds.cern.ch/record/1407893}{LHCb-PUB-2011-026;
  CERN-LHCb-PUB-2011-026}, CERN, Geneva, 2012\relax
\mciteBstWouldAddEndPuncttrue
\mciteSetBstMidEndSepPunct{\mcitedefaultmidpunct}
{\mcitedefaultendpunct}{\mcitedefaultseppunct}\relax
\EndOfBibitem
\bibitem{HULSBERGEN2009471}
W.~D. Hulsbergen, \ifthenelse{\boolean{articletitles}}{\emph{{The global
  covariance matrix of tracks fitted with a Kalman filter and an application in
  detector alignment}},
  }{}\href{https://doi.org/https://doi.org/10.1016/j.nima.2008.11.094}{Nucl.\
  Instrum.\ Meth.\  \textbf{A600} (2009) 471}\relax
\mciteBstWouldAddEndPuncttrue
\mciteSetBstMidEndSepPunct{\mcitedefaultmidpunct}
{\mcitedefaultendpunct}{\mcitedefaultseppunct}\relax
\EndOfBibitem
\bibitem{VIRET2008157}
S.~Viret, C.~Parkes, and M.~Gersabeck,
  \ifthenelse{\boolean{articletitles}}{\emph{{Alignment procedure of the LHCb
  vertex detector}},
  }{}\href{https://doi.org/https://doi.org/10.1016/j.nima.2008.07.153}{Nucl.\
  Instrum.\ Meth.\  \textbf{A596} (2008) 157}\relax
\mciteBstWouldAddEndPuncttrue
\mciteSetBstMidEndSepPunct{\mcitedefaultmidpunct}
{\mcitedefaultendpunct}{\mcitedefaultseppunct}\relax
\EndOfBibitem
\end{mcitethebibliography}
 
\newpage
\centerline
{\large\bf LHCb collaboration}
\begin
{flushleft}
\small
R.~Aaij$^{32}$,
A.S.W.~Abdelmotteleb$^{56}$,
C.~Abell{\'a}n~Beteta$^{50}$,
T.~Ackernley$^{60}$,
B.~Adeva$^{46}$,
M.~Adinolfi$^{54}$,
H.~Afsharnia$^{9}$,
C.~Agapopoulou$^{13}$,
C.A.~Aidala$^{86}$,
S.~Aiola$^{25}$
% ...

{\footnotesize \it

% ...
$^{9}$Universit{\'e} Clermont Auvergne, CNRS/IN2P3, LPC, Clermont-Ferrand, France\\
% ...
$^{13}$LPNHE, Sorbonne Universit{\'e}, Paris Diderot Sorbonne Paris Cit{\'e}, CNRS/IN2P3, Paris, France\\
% ...
$^{25}$INFN Sezione di Milano, Milano, Italy\\
% ...
$^{32}$Nikhef National Institute for Subatomic Physics, Amsterdam, Netherlands\\
% ...
$^{46}$Instituto Galego de F{\'\i}sica de Altas Enerx{\'\i}as (IGFAE), Universidade de Santiago de Compostela, Santiago de Compostela, Spain\\
% ...
$^{50}$Physik-Institut, Universit{\"a}t Z{\"u}rich, Z{\"u}rich, Switzerland\\
% ...
$^{54}$H.H. Wills Physics Laboratory, University of Bristol, Bristol, United Kingdom\\
% ...
$^{56}$Department of Physics, University of Warwick, Coventry, United Kingdom\\
% ...
$^{60}$Oliver Lodge Laboratory, University of Liverpool, Liverpool, United Kingdom\\
% ...
$^{68}$Syracuse University, Syracuse, NY, United States\\
% ...
$^{86}$University of Michigan, Ann Arbor, United States, associated to $^{68}$\\
\bigskip
% ...
\medskip
}
\end{flushleft}

\end{document}